# Roadmap on Advances in Visual and Physiological Optics

**Jesús E. Gómez-Correa[1,28,*], Brian Vohnsen[2,28], Barbara K. Pierścionek[3,28], Sabino Chávez-Cerda[1,28], Sabine Kling[4], Jos J. Rozema[5,6], Raymond A. Applegate[7], Giuliano Scarcelli[8], J. Bradley Randleman[9], Alexander V. Goncharov[10], Amy Fitzpatrick[2], Jessica I. W. Morgan[11,12], Austin Roorda[13], David A. Atchison[14], Juan P. Trevino[15,16], Alejandra Consejo[17], Charlie Börjeson[18], Linda Lundström[18], Seung Pil Bang[19], Geunyoung Yoon[7], Karol Karnowski[20,21], Bartlomiej J. Kaluzny[22], Ireneusz Grulkowski[23], Sergio Barbero[24], Pablo Artal[25], Juan Tabernero[25], Pete Kollbaum[26], and Stéphanie C. Thébault[27]**

[1]Instituto Nacional de Astrofísica, Óptica y Electrónica, Coordinación de Óptica, Pue. 72840, Mexico
[2]Optics Group, School of Physics, University College Dublin, Beech Hill, Ireland
[3]Faculty of Health, Medicine and Social Care, Medical Technology Research Centre, Anglia Ruskin University, Bishops Hall Lane, Chelmsford, United Kingdom
[4]ARTORG Center for Biomedical Engineering Research, University of Bern, Freiburgstrasse 3, 3010 Bern, Switzerlandfir
[5]Visual Optics Lab Antwerp, Faculty of Medicine and Health Sciences, University of Antwerp, Wilrijk, Belgium
[6]Dept. of Ophthalmology, Antwerp University Hospital, Edegem, Belgium
[7]College of Optometry, University of Houston, Houston TX, USA
[8]University of Maryland, College Park MD
[9]Cole Eye Institute, Cleveland OH
[10]School of Natural Sciences, University of Galway, Ireland
[11]Scheie Eye Institute, Department of Ophthalmology, University of Pennsylvania, Philadelphia, PA, USA
[12]Center for Advanced Retinal and Ocular Therapeutics, University of Pennsylvania, Philadelphia, PA, USA
[13]Herbert Wertheim School of Optom. Vis. Sci., University of California, Berkeley, Berkeley CA 94720-2020
[14] Centre for Vision and Eye Research, Queensland University of Technology, Kelvin Grove, Australia
[15]Universidad Politécnica de Puebla, Cuanalá Puebla 72640, Mexico
[16]Tecnológico de Monterrey, México
[17]Aragon Institute for Engineering Research (I3A), University of Zaragoza, Zaragoza, Spain
[18]KTH Royal Institute of Technology, Department of Applied Physics, Stockholm
[19]Department of Ophthalmology, Keimyung University Dongsan Medical Center, Daegu, South Korea
[20]International Centre for Translational Eye Research, ul. Skierniewicka 10A, 01-230 Warsaw, Poland
[21]Institute of Physical Chemistry, Polish Academy of Sciences, ul. M. Kasprzaka 44/52, 01-224 Warszawa, Poland
[22]Department of Ophthalmology, Collegium Medicum, Nicolaus Copernicus University, ul. M. Curie Skłodowskiej 9, 85-094 Bydgoszcz, Poland
[23]Institute of Physics, Faculty of Physics, Astronomy and Informatics, Nicolaus Copernicus University, ul. Grudziadzka 5, 87-100 Toruń, Poland
[24]Instituto de Óptica (CSIC), Serrano 121, Madrid, Spain
[25]Laboratorio de Optica, Universidad de Murcia, 30100 Murcia, Spain
[26]Indiana University, 800 East Atwater Avenue, Bloomington, IN 47405, USA
[27]Laboratorio de Investigación Traslacional en Salud Visual, Instituto de Neurobiología, Universidad Nacional Autónoma de México (UNAM), Campus Juriquilla, Querétaro, Mexico.
[28]Guest editors of the Roadmap.
* Corresponding Author: jgomez@inaoep.mx

## Abstract

The field of visual and physiological optics is undergoing continuous significant advancements, driven by a deeper understanding of the human visual system and the development of cutting-edge optical technologies. This Roadmap, authored by leading experts, delves into critical areas such as corneal biomechanical properties, keratoconus, and advancements in corneal imaging and elastography. It explores the intricate structure-function relationship within the eye lens, offering new perspectives through lens models and ray tracing techniques. The document also covers advancements in retinal imaging, highlighting the current state and future directions, and the role of adaptive optics in evaluating retinal structure and function in both healthy and diseased eyes. Furthermore, it addresses the modelling of ocular surfaces using different mathematical functions and examines the factors affecting peripheral image quality in the human eye, emphasizing the importance of these aspects in visual performance. Additional topics include schematic and functional models of the human eye, the impact of optical and chromatic aberrations, and the design of contact, and intraocular lenses. Finally, the Roadmap addresses the intersection of neurosciences with vision health, presenting a comprehensive overview of current research and future trends aimed at improving visual health and optical performance. Ultimately, this Roadmap aims to serve as a valuable resource for ophthalmologists, optometrists, vision scientists, and engineers dedicated to advancing the field of visual and physiological optics.

Contents

# 1. Introduction

Jesús E. Gómez-Correa[1], Brian Vohnsen[2], Barbara K. Pierścionek[3], and Sabino Chávez-Cerda[1]

[1]Instituto Nacional de Astrofísica, Óptica y Electrónica, Coordinación de Óptica, Pue. 72840, Mexico
[2]Optics Group, School of Physics, University College Dublin, Beech Hill, Ireland
[3]Faculty of Health, Medicine and Social Care, Medical Technology Research Centre, Anglia Ruskin University, Bishops Hall Lane, Chelmsford, United Kingdom

[ jgomez@inaoep.mx; brian.vohnsen@ucd.ie; barbara.pierscionek@aru.ac.uk; sabino@inaoep.mx ]

From ancient Greece to Kepler's revolutionary ideas, vision was interpreted as an interaction between the eye and external objects with differing views on the source of optical elements. For example, by the mid-first millennium B.C., the first documented speculations about vision had emerged. In popular Greek belief, it was thought that the eye emitted a fire whose rays explored the surface of the observed object, known as "visual fire." However, not all philosophers accepted this theory, most notably—the atomists— who opposed it and reduced all sensations to the impact of atoms emanating from the object onto the organ of vision. Other philosophers believed that vision depended on a medium that propagated the form, which was considered inert and sensitive. Others believed in the idea of rays that emanated from the eye at defined angles to observe objects, a concept that formed the basis of Geometric Optics. In the 11th century, Alhazen revolutionized the theory of vision by proposing that vision depended on light entering the eye after being diffusely reflected by the surface of objects. Centuries later, Kepler completed Alhazen's idea by arguing that the image of the object was formed on the retina. He understood that this image was generated by the intersection of all rays coming from a luminous point, allowing the distance to the object to be measured. With this idea, Kepler turned the eye into an optical instrument and demonstrated how instruments could correct or enhance vision [1,2].

Today, the field of optics has evolved far beyond its early theoretical foundations. Modern advancements in optical technologies, such as lasers, imaging systems, optical sensors, adaptive optics in combination with computational modelling, have all been vital in helping us understand how the biology of the eye is linked to its optical function and the visual system. These innovations have led to the development of sophisticated instruments and techniques for diagnosing and treating eye conditions and exploring the intricate workings of the human eye at a level of detail unimaginable in earlier times.

Building on these advancements, the "Roadmap on Advances in Visual and Physiological Optics," authored by leading experts, delves into critical areas that are shaping the future of vision science. This comprehensive document explores topics from the cornea and all the way through to the brain and visual cortex including corneal biomechanics, keratoconus, and cutting-edge techniques in corneal imaging and elastography. It also examines the complex structure-function relationship within the eye lens, utilizing advanced lens models and ray tracing techniques to offer new insights. The Roadmap highlights the latest advancements in retinal imaging, adaptive optics, and the modelling of ocular surfaces, addressing their implications for visual performance and the treatment of eye diseases. Additionally, it covers the impact of optical and chromatic aberrations, the design of progressive lenses, contact lenses, and intraocular lenses, as well as the integration of the retina and neurosciences into vision health. By providing a forward-looking overview of current research and emerging trends, we hope that this Roadmap will serve as an invaluable resource for professionals dedicated to advancing the field of visual and physiological optics, guiding future innovations aimed at enhancing both optical performance and overall visual health of humanity.

## 2. Corneal biomechanical properties

Sabine Kling

ARTORG Center for Biomedical Engineering Research, University of Bern, Freiburgstrasse 3, 3010 Bern, Switzerland

[ sabine.kling@unibe.ch ]

**Status**

Understanding the mechanical behavior of the cornea is critical to discern health from disease, but also to accurately predict surgical outcomes. The cornea exhibits complex mechanical properties, which allow for its transparency, dome-shaped geometry and its extraordinary stable refractive performance, even under varying intraocular pressure (IOP). Recent advances in this field include higher resolved [1,2] and non-contact [3,4] mechanical characterization techniques, computational modeling [5,6], and new clinical applications [7,8], enhancing our knowledge of corneal biomechanics and its role in ocular pathologies.

Keratoconus is a progressive ectatic disorder that locally degrades the tissue's mechanical integrity resulting in corneal thinning, localized steepening and severe visual degradation. New diagnostic tools aim at detecting subclinical keratoconus, facilitating early diagnosis and intervention. Myopia, another common refractive condition, has recently been investigated in the context of corneal biomechanics. Myopic eyes, particularly those with severe elongation, were reported to exhibit altered biomechanical properties, when measured in vivo [9-11]. However, this observation might be a measurement artefact, resulting from the use of evaluation metrics outside of their calibrated range (axial eye length). More objective evaluations of the role of corneal biomechanics in myopia will be necessary, before this relationship can be used to assess the progression of myopia and the associated risk of complications.

The most often conduced mechanical in vivo measurements of the cornea assess the deformation dynamics induced by an air-puff via the Ocular Response Analyzer [4] (ORA) and the Corvis ST [3] (CST). While the originally mostly geometrical parameters retrieved from air-puff measurements were strongly dependent on confounding factors such as IOP and ocular geometry (corneal thickness), more recently advanced parameters have been derived based on large clinical data sets to define more robust parameters (Corvis biomechanical index [12] – CBI, tomographic biomechanical index [13] – TBI) to distinguish normal from keratoconus and ectatic corneas. Further air-puff parameters are the biomechanically-corrected IOP (bIOP) providing a less biased measure of the actual IOP and the stress-strain index [3] (SSI) objectively assessing the corneal stiffness independent of corneal thickness and IOP. Experimental techniques such as micro-forced shear-wave [2], reverberant [14]) or quasi-static[15] optical coherence elastography (OCE) pose promising alternatives to quantify corneal biomechanics with higher spatial resolution providing a more comprehensive assessment of tissue heterogeneity. Another promising approach is motion-tracking Brillouin microscopy, which successfully quantified the local decrease in corneal stiffness in keratoconus patients [1].

Advances in finite element analysis reduce the amount of approximations by accounting for the pre-stress by the IOP, collagen fibril orientation, non-linear material properties and patient-specific geometries with the aim of predicting and optimizing treatment protocols, but also inversely retrieving corneal material properties from in vivo measurements [16].

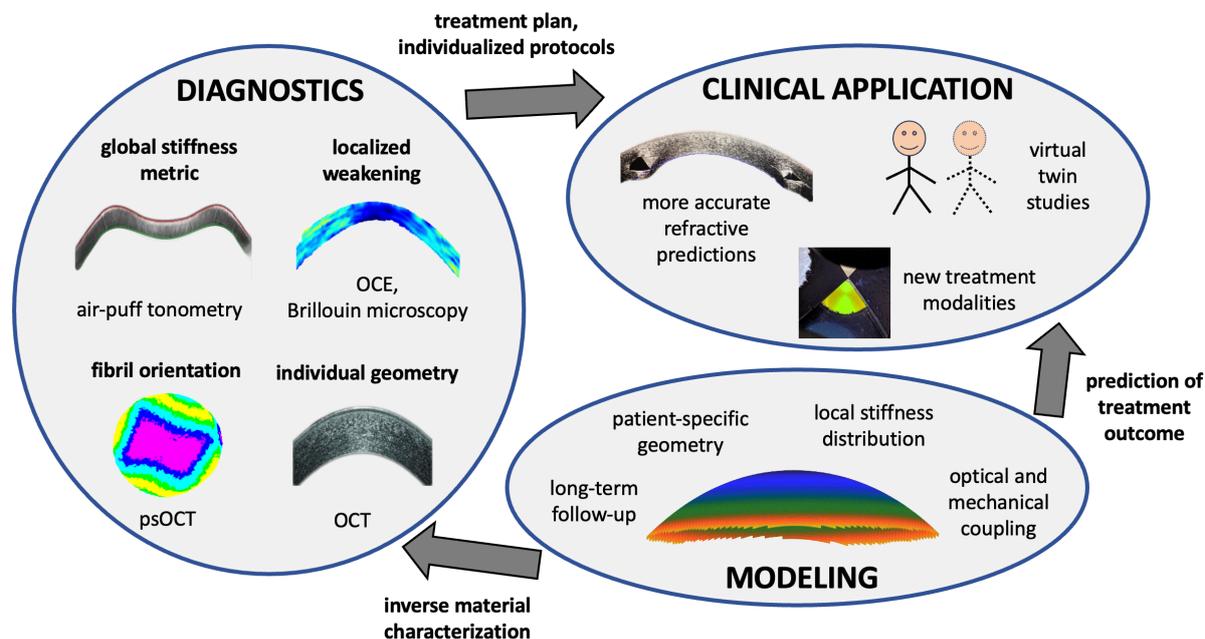

**Figure 1.** Key areas of corneal biomechanics and their interactions.

**Current and Future Challenges**

A primary challenge in current clinical biomechanical assessments is their indirect nature and the lack of standardized measurement techniques. Differences in the principles and outputs of ORA and CST prevent a comparison of their mechanical measures. Their strong dependency on intraocular pressure and corneal thickness demand for complex processing algorithms to derive corneal stiffness, which themselves rely on certain approximations on ocular geometry, material properties and boundary conditions, preventing their application as a universal measurement tool in all patients.

The complex layered structure of the cornea and its anisotropic, visco-hyperelastic material properties make it difficult to model and analyze accurately. Variations in collagen fiber orientation and corneal hydration levels can only be indirectly estimated in vivo and the assessment of regional differences in corneal stiffness via elastography rests in its infancy – making important modelling parameters difficult to obtain. In consequence, models and diagnostic tools often simplify these aspects, limiting their accuracy and applicability in clinical practice.

Significant inter-individual variability in corneal biomechanics exists due to genetic, environmental and lifestyle factors. This variability makes generalized diagnostic and treatment protocols unreliable and demands for personalized assessment tools and treatment protocols. The integration of both, biomechanical and other clinically available diagnostic information, such as topography, tomography and optical coherence tomography (OCT) in more sophisticated models promise more refined diagnostics and more accurate predictions of the surgical outcome. This is particularly relevant for corneal surgeries, such as LASIK, SMILE, and corneal cross-linking, which alter the corneal structure and mechanics.

A more detailed understanding of the factors influencing corneal biomechanics could also allow a better patient-specific risk estimation of developing iatrogenic ectasia and supplement pre-surgical biomechanical screening before laser refractive surgeries. Moreover, being able to model the dynamics of degenerative corneal diseases could help forecasting the individual speed of progression and assist in determining the adequate treatment schedule and protocol.

A still unmet and unaddressed challenge is the prediction of post-surgical wound healing effects, which are inherently accompanied by biomechanical changes. This includes epithelial and stromal remodeling after intracorneal ring segment implantation and laser refractive surgery, but also ongoing corneal flattening after corneal cross-linking treatment. Currently, these processes limit the predictability of long-term post-surgical refractive results.

**Advances in Science and Technology to Meet Challenges**

Future advancements in imaging and measurement technologies are expected to enhance the resolution of biomechanical assessments and their sensitivity to detect early disease stages. Techniques such as optical coherence elastography [2,15] (OCE) could provide more detailed insights into corneal biomechanics at both the macroscopic and microscopic levels. With advances in imaging speed, refractive and biomechanical treatment monitoring during CXL treatment might become possible and allow tailored biomechanical modification in an individual patient. Polarization sensitive OCT [17] is a promising tool to provide information about the patient-specific fiber orientation in vivo, providing a relevant modeling input. In the same line, advanced OCT signal interpretation algorithms such as nanosensitive OCT [18] demonstrated their potential to extract ultrastructural information out of standard OCT scans and could similarly provide insights into the collagen network.

The development of more sophisticated computational models that can simulate the complex mechanical interaction of the cornea and other ocular tissues (crystalline lens, sclera) and directly relate mechanical with optical changes is a major future challenge. Comprehensive opto-biomechanical eye models [19,20] are being developed promising the evaluation of distinct surgical interventions in a digital twin before choosing the best treatment option. Enhancing these models by individual opto-mechanical measures and validating their predictions under distinct optical and mechanical demands will require significant advancements in computational power and algorithms.

The application of artificial intelligence might allow to reduce the computational load of numerical simulations and to identify novel biomarkers for corneal degradation in the future. Big data analytics could also provide deeper insights into the progression of diseases like keratoconus and myopia, as well as the long-term stability of surgical interventions. With digital patient files readily available and decreasing storage costs, the collection of extensive datasets over time has become more feasible.

**Concluding Remarks**

Corneal biomechanics is a dynamic and critical area of ophthalmic research. Advances in diagnostic tools – like the ORA and CST, along with innovative techniques such as OCE and Brillouin microscopy – have significantly enhanced our ability to assess corneal biomechanics in vivo. These developments are vital for the early detection and management of degenerative conditions like keratoconus and myopia, where biomechanical properties play an important role in disease progression.

Current challenges include standardizing measurement techniques, integrating biomechanical data with other diagnostic modalities, and accounting for inter-individual variability. Future challenges lie in developing more advanced imaging technologies, improving computational models, and implementing personalized medicine approaches. Addressing these challenges requires ongoing interdisciplinary research, technological innovation, and clinical collaboration. As we continue to refine our understanding and tools, the potential for improving patient outcomes through better diagnostic and treatment strategies in corneal biomechanics is substantial.

**Acknowledgements**

This work received funding from the European Union's HORIZON 2020 research and innovation programme under grant agreement No. 956720.

## 3. Keratoconus, corneal aberrations, and their correction

Jos J. Rozema[1,2] and Raymond A. Applegate[3]

[1]Visual Optics Lab Antwerp, Faculty of Medicine and Health Sciences, University of Antwerp, Wilrijk, Belgium
[2]Dept. of Ophthalmology, Antwerp University Hospital, Edegem, Belgium
[3]College of Optometry, University of Houston, Houston TX, USA

[ jos.rozema@uantwerpen.be ; raappleg@central.uh.edu ]

**Status**

Keratoconus is progressive corneal ectasia that typically first manifests in the late teenage years with an estimated prevalence [1] of about *1:375* and a male to female ratio of *2.6:1* [2]. The condition is bilateral but does not necessarily affect both eyes simultaneously or to the same degree, often leading to very large differences in visual quality between eyes of the same individual, while the rate of progression and severity can vary widely [3]. Although the cause is unknown, there is a hereditary component, while eye rubbing and allergies are known to exacerbate it [2]. The irregular corneal shape found in keratoconus, pellucid degeneration, corneal trauma, or after some corneal surgical interventions, lead to increased amounts of higher order wavefront aberrations. Often, such aberrations lead to a decrease in acuity. Whether high contrast acuity is lost or not, visual image quality is decreased. As the onset of the disease occurs just before early adulthood, when individuals head off to college or establish their careers and families, it is a particularly difficult time for them to learn about having an eye disease that will progressively affect their vision. Based on their visual experiences before the onset of the disease patients expect solutions that return their original, healthy sight. But while challenging, many solutions exist that help patients regain some of their lost visual image quality. Wavefront guided scleral lenses, in many cases, reduce aberrations to age and pupil size normal levels.

**Current and Future Challenges**

Much like typical eyes, individuals with early keratoconus may be satisfied with the visual image quality provided by standard spherical or sphero-cylindrical spectacles, or soft contact lens corrections. As the disease progresses, however, the increasing amounts of higher order aberrations will cause such corrections to gradually fail to meet the patients' visual needs. The next level of correction is the corneal gas permeable **rigid contact lens** that, if designed and fitted correctly, provide a smooth first surface while creating an index matching tear lens between the lens and cornea. This significantly reduces the defocus and astigmatism, as well as the higher order aberrations by approximately *60%* [4]. More advanced cases see the clearing between cornea and lens become so narrow that both may start to rub each other, leading corneal scarring or ulcers. To prevent scarring or ulcers, the standard of care becomes **scleral lenses** that vault the distorted corneal surface. But as the disease progresses the index matching of rigid lenses may no longer be adequate to the patient's needs as these only partially compensate the higher order aberrations of the anterior corneal surface ignore those of the posterior corneal surface [5]. These limitations are beginning to be addressed by **WaveFront Guided (WFG) rigid scleral lenses**, designed to return the highly aberrated eyes to normal levels of wavefront error (WFE) [4]. These lenses require a highly precise registration with the underlying WFE, involving both rotation and translation, where the tolerance decreases with an increasing aberration severity (e.g., for highly aberrated eyes the WFG lens may only shift less than *0.2 mm* or rotate less than *5°* to retain above normal levels of visual image quality) [5]. This requires an exceptional stability that only few current lenses can provide.

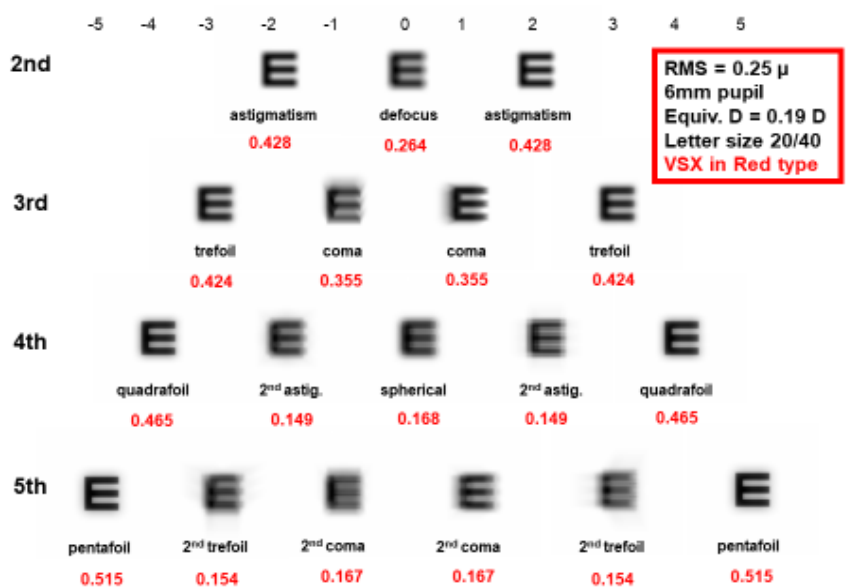

**Figure 1**. A 20/40 high contrast letter E aberrated by 0.25 μm of RMS wavefront error over a 6 mm diameter pupil for each individual Zernike aberration of the 2nd through 5th radial order. For defocus, 0.25 μm over a 6mm pupil is equivalent to 0.19 D. Numbers in the red are values for the visual Strehl ratio (VSX) calculated in the spatial domain, varying between 0 (worst) and 1 (best). Notice that even though RMS WFE is the same for each simulation, image quality is different as also reflected in the VSX values.

Besides refractive corrections, there are also surgical solutions to keratoconus. These include **corneal crosslinking**, a widely used therapy that aims to slow down or even halt progression by increasing the corneal stiffness. It can provide stability for up to *15* years [6] and can be reapplied if needed. Although crosslinking cannot eliminate the large corneal deformations, it may occasionally flatten the cornea [7]. This flattening may be enhanced by customizing the treatment pattern to the patient's corneal topography or corneal stiffness maps [8], but this option is not yet widely available and in the near future unlikely to return the cornea to normal optical function. Other surgical approaches attempt to smooth the anterior corneal surface by implanting **corneal ring segments** made of a polymer or an allogenic corneal material, or in more severe cases by a **transplanting** either the anterior corneal layers of the cornea or the entire cornea. Although surgical techniques can improve visual performance of keratoconic eyes, they rarely if ever provide age-appropriate levels of visual image quality [9] with a sphero-cylindrical correction.

**Advances in Science and Technology to Meet Challenges**

To optimize the WFG corrections, it is crucial to follow the ANSI guidelines [10] of using a fixed coordinate system with an origin at the pupil centre for WFE measurements, aberrometer alignment, and scleral lens positioning over time and during manipulation. Moreover, to properly track eye rotation, the fixation point must be placed on the aberrometer's optical axis, which must pass through the centre of the dilated pupil while the patient looks at a fixation target [11]. These are current guidelines, but not always followed to the letter in current ocular aberrometers.
Next, it is essential to use metrics to assess the visual image quality, as the widely used total and higher order root-mean-square (RMS) WFE are not appropriate for this purpose for three reasons: (1)

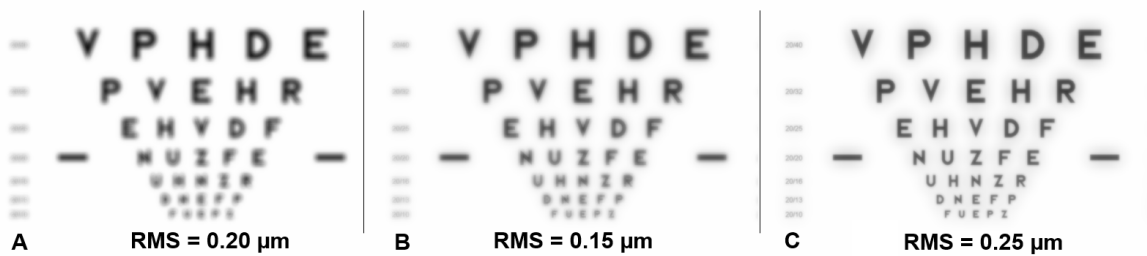

**Figure 2.** The effect of 0.20 µm RMS of defocus (A), 0.15 µm RMS of spherical aberration (B) and the combination of the two producing $\sqrt{0.20^2 + 0.15^2} = 0.25\ \mu m$ of RMS (C) on a high contrast logMAR acuity chart. Notice that the image quality is worse in panels A and B than in C, despite having a lower RMS.

different aberrations with the same RMS impact vision differently (Figure 1)[12]; (2) aberrations interact with each other to improve or decrease visual performance (Figure 2)[13]; end (3) RMS only has meaning for the pupil diameter at which it was measured. To address this, alternatives metrics such as the visual Strehl ratio (*VSX*) were developed [14], which considers the neural processing and provides meaningful values with interesting attributes [9, 15, 16], while being independent of pupil size or RMS [16]. Regardless, **new visual image quality metrics** should be developed to better predict the various aspects of visual performance beyond high contrast visual acuity.

These considerations demonstrate that the goal of WFG corrections on a highly aberrated eye should not be to drive the RMS WFE to zero, which is technically impossible and would take the image resolution beyond the sampling capacity of the foveolar cones [17]. Instead, one should optimize the visual image quality of that eye under the best sphero-cylindrical correction to match the eye's neural processing capabilities for its largest physiological pupil diameter. This is done by bringing the *VSX* to within one standard deviation of the age-normal reference values of healthy eyes [5]. Setting clear goals for the correction and specifying a clear metric by which to assess its effectiveness will help avoid arbitrary approaches to WFG lenses. Furthermore, clear **standards** for the development and manufacturing should be drafted and updated on a regular basis to make WFG lenses a **safe and affordable option** to patients, while clinicians should be trained to the unique challenges of WFG lenses and the use of WFE data to show their patients the origins of their complaints.

Looking at **customized crosslinking**, the therapeutic may be improved by developing methods to **map corneal weak spots** using e.g. optical coherence elastography and subsequently target those areas for treatment. Meanwhile, biomechanical models can help develop **new treatment patterns** to optimize the corneal flattening to facilitate sphero-cylindrical contact lens corrections and optimize long-term stability, while other algorithms could assess the **risk of keratoconus progression** to identify those in need of crosslinking in the short term.

Alternative forms of correction could also be envisioned, such as a sphero-cylindrical contact lens that **masks** the regions with the highest WFE, WFG intraocular lenses or **adaptive optics spectacles** that adapt to wherever the eye is looking.

**Concluding Remarks**

The continued development of two currently existing technologies has the best chance to better address the visual needs of people with keratoconus. The first, customized crosslinking combined with corneal elasticity mapping, will help stop the progression of the disease and perhaps reverse the corneal deformation to a minor degree. The second technology is, depending on the visual image quality, either sphero-cylindrical or wavefront guided (scleral) contact lenses to return the visual image quality to age-appropriate normal levels. To ensure WFG lenses find widespread adoption for patients that need them, it is essential to create standards for both manufacturers and clear and verified guidelines for clinicians to ensure that WFG lenses become a safe and affordable solution for patients to meets their visual needs.

## 4. Advancements in corneal imaging and elastography

Giuliano Scarcelli[1] and J. Bradley Randleman[2],
[1]University of Maryland, College Park MD
[2]Cole Eye Institute, Cleveland OH

[ scarc@umd.edu; RANDLEJ@ccf.org ]

**Status**

The cornea is the primary determinant of visual acuity, providing 70% of the eye's total refractive power. Corneal power is determined by corneal morphology, which in the healthy eye exists in relative equilibrium within the mechanical balance between atmospheric pressure and intraocular pressure (IOP) due to the stroma's high collagen content and transverse lamellar collagen fiber structure. It is now widely recognized, from first principles, computational studies, and experimental measurements, that biomechanics drive corneal shape and thus visual acuity. [1-3]

Corneal biomechanics impact two intertwined issues of major clinical importance: keratoconus and myopia. Keratoconus, a corneal disease characterized by progressive corneal warpage, results in vision loss, severely impacts quality of life, and its prevalence (2-5%) is significantly higher than thought.[4] Myopia impacts ~1.4 billion people worldwide, with prevalence expected to double, affecting more than 50% of the US population, by 2050.[5] The economic burden of uncorrected refractive error is already substantial, with estimates exceeding $200 billion in productivity lost from >150 million persons with uncorrected refractive error. [6]

Keratoconus progression can be stopped with corneal cross-linking (CXL) with minimal visual loss if detected early; however, insurance companies require morphologic proof of disease progression across sequential visits before authorizing treatment. CXL protocols offer modest improvement in visual acuity and impart risk for sight-threatening complications. Thus, early and accurate KC progression detection provides the greatest benefit to patients before vision is irreversibly compromised.

Laser vision correction (LVC) procedures (eg. LASIK) effectively treat myopia, compares favourably to contact lens wear, and could help address the global burden of myopia; yet less than 10% of eligible patients undergo LVC, with most citing safety concerns as the major factor.[7, 8] Refractive surgery screening remains inherently conservative, as only morphological evaluations are currently available in the clinic; thus, many otherwise reasonable candidates are excluded from surgery due to the risk of developing postoperative ectasia, which continues to occur despite the utilization of multiple screening strategies including artificial intelligence approaches. [9]

For both keratoconus and myopia, these challenges exist because, to date, despite major efforts to develop techniques to measure local corneal mechanical properties, current technologies have demonstrated low sensitivity to detect clinically-relevant features of ectasia susceptibility. The lack of effective biomechanical measurements has forced clinicians to rely on morphologic surrogates, e.g. curvature and thickness, which are insufficient to optimally identify keratoconus before vision is compromised, screen at-risk surgical candidates, [10] or predict treatment outcomes after LVC or corneal cross-linking (CXL). Thus, the need to accurately identify the underlying corneal biomechanical profile that signifies corneal ectasia progression risk has never been greater.

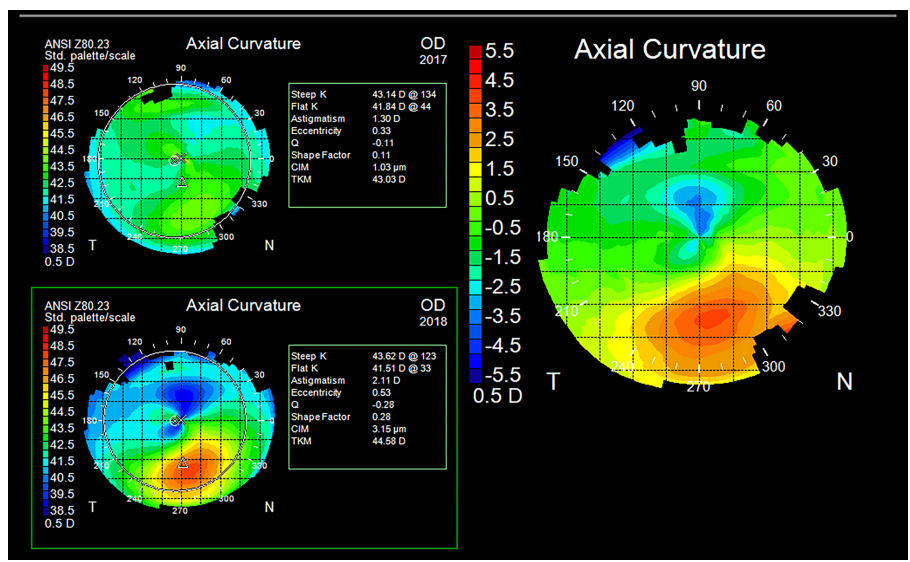

**Figure 1.** Placido imaging of a patient without clear evidence of keratoconus in the right eye at their initial presentation in 2017 (upper left image) who suffered disease progression over the course of one year (bottom left image), with focal steepening of 4D and loss of visual acuity.

**Current and Future Challenges**

Intense efforts have been put forward to develop technology that measures corneal biomechanics in vivo. The only commercial device approved by the FDA for corneal biomechanical measurements, the Ocular Response Analyzer (ORA, Reichert, Inc.), quantifies one dimensional corneal deformation and recovery dynamics during corneal deformation by air puff, using a parameter called corneal hysteresis (CH). CH is a difference between the ingoing and outgoing pressures when the cornea becomes flat (at applanation) and represents viscous damping in the anterior segment. ORA metrics are reduced in keratoconus in proportion to the keratoconus severity. However, their sensitivity and specificity are insufficient to differentiate early keratoconus from normal corneas. [11]
The CorvisST (Oculus, Inc.) combines air-puff induced deformation and a high-speed Scheimpflug camera to measure the mechanical response of the cornea. Dynamic optical coherence tomography (OCT) coupled with air-puff has also been evaluated. All of these techniques require sophisticated image analysis and numerical computation based on finite element analysis to solve an inverse problem that accounts for many variables beyond corneal elasticity. In addition, they are only sensitive to the global stiffness of the cornea and are unable to identify localized focal weakening. [12]
Ultrasound techniques take advantage of different phenomena from air-puff mediated image analysis and thus allows local measurement of the elastic modulus of the tissue (at MHz frequency). However, ultrasound requires the excitation of relatively high-energy ultrasound via acoustic transducer and coupling gel. Although shear wave imaging by ultrasound is promising, further development and validations in vivo will be necessary to assess the potential of these techniques. [13]

**Advances in Science and Technology to Meet Challenges**

Two optical elastography techniques have emerged in the research realm to address the challenges of this field: optical coherence elastography (OCE) and Brillouin microscopy.[14, 15] Optical coherence elastography takes advantage of the relationship between tissue stiffness and the speed of mechanical waves propagating in the tissue. The mechanical waves are usually launched via acoustic radiation force applied either via coupling gel or air-coupled; the speed of the wave is measured via Optical Coherence Tomography. Several ex vivo studies and in vivo studies in animals have demonstrated the potential of the technique; translation of this technology in vivo to the ophthalmology clinic is currently ongoing. [16]

Brillouin microscopy is based on the interaction of incident light and thermal acoustic phonons in a sample. As a result of this interaction, the scattered light has a slightly different wavelength so that by measuring the difference in wavelength between incident and scattered light, one can extract the local longitudinal modulus of the tissue. Early prototypes of clinical instruments based on Brillouin microscopy demonstrated the ability to distinguish normal vs ectatic corneas. [17] However, Brillouin instruments based on early designs suffered from motion artifacts that hindered sensitivity for clinically-useful tasks such as early keratoconus detection or the differentiation of LVC mechanical impact. [18]

Recently, the next generation of Brillouin microscopes features co-located Optical coherence tomography to enable correction of motion artifacts.[19] The resulting high sensitivity provided the first demonstration of focal weakening in early and subclinical keratoconus in *vivo*. [20, 21] Clear differences in regional Brillouin values were observed between the groups as demonstrated by significant differences for controls and both subclinical and early KC in several spatially resolved metrics with receiver operating characteristic (ROC) curves reaching AUROC=1. In early KC cases, Brillouin shift values performed better traditional metrics such as maximum curvature (Kmax) and thinnest pachymetry. For eyes with subclinical keratoconus that were not detectable using morphologic analysis via Scheimpflug tomography, motion-tracking Brillouin microscopy showed a clear focal weakening with a statistically significant reduction in minimum Brillouin shift. This is the first experimental evidence demonstrating metrics quantifying focal weakening in keratoconus corneas that can reach or surpass clinical/morphological metrics.

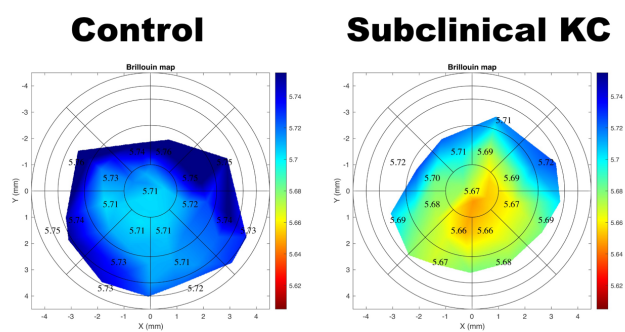

**Figure 2.** Motion tracking Brillouin imaging showing regional differences between a normal cornea (Control) and an eye with subclinical keratoconus. MT Brillouin shift values are uniform across the normal control eye, while there is clear focal weakening in evident in the subclinical KC eye.

**Concluding Remarks**

The field of corneal imaging and elastography is at a pivotal point; after decades where it has been recognized that biomechanics drive corneal shape and thus visual acuity, finally sensitive technologies to leverage mechanical measurements are coming into the clinic. At this stage, the first milestone has been demonstrated, i.e. the superiority of sensitive mechanical metrics over morphological ones for corneal diagnosis and management. The future needs a new leap in technology development to make elastography technologies user friendly, cost-effective, commercially available to facilitate wide adoption in clinical settings.

**Acknowledgements**

This work was supported in part by NIH grants R01 EY028666, R01EY032537.

## 5. Structure and function relationship in the eye lens

Barbara K. Pierścionek

Faculty of Health, Medicine and Social Care, Medical Technology Research Centre, Anglia Ruskin University, Bishops Hall Lane, Chelmsford, United Kingdom

[ barbara.pierscionek@aru.ac.uk ]

**Status**

A third of the refractive power of the human eye for distance vision is provided by the eye lens. The lens is also able to change its shape, or accommodate, to enable the eye to focus over a range of distances. Accommodative capacity gradually decreases with age until the sixth decade of life whereupon the human eye is unable to focus on near objects. The lens has a lamellar structure with new layers of long thin fibre cells growing over existing tissue. There is no concomitant cellular loss which creates a structure that has the oldest cells in the centre and the youngest in the periphery. The distribution and concentration of proteins in the various layers of the lens varies increasing from the periphery of the lens to its centre and hence creating a gradient of protein concentration which manifests as a gradient of refractive index [1]. This gradient index contributes to the superior degree of image quality in the lens. The changes in optical quality and accommodative capacity and their reduction with age are caused by alterations in the lens proteins. To date, these critical structure/function links are not understood and yet they are fundamental to an enhanced understanding and improved treatment of both presbyopia (loss of accommodative ability) and cataract (loss of optical function).

The major proteins in the lens are the crystallins [2] and these constitute around 35% of the lens wet weight. They are categorised based on biophysical and chemical properties and in the human lens are broadly identified as $\alpha, \beta$ and $\gamma$-crystallins [2]. These proteins are synthesised in the lens fibre cells and there is no loss of proteins throughout life. Hence, the ages of the proteins vary across the lens with the oldest proteins in the centre and the youngest in the peripheral layers [1]. Changes in the proteins with time are thought to lead to alterations in their confirmational structure and in their relationship with water, leading to more free water (unbound to protein) with age [3]. Some of these changes are thought to lead to protein aggregation and when this is sufficiently advanced, vision is impaired because of scatter and/or absorption of light, the lens loses transparency and the clinical diagnosis is cataract.

A further group of lenticular proteins that are much less common than the crystallins are the aquaporins, which are found in lens cell membranes and are responsible for transport of nutrients and water to the lens fibre cells. The function of these proteins is vital for a healthy microenvironment and for maintaining transparency.

**Current and Future Challenges**

Although it is accepted that the distribution and concentration of the different crystallin proteins creates the refractive index and that the aquaporins, or water channel proteins, play a role in water and nutrient transport, how changes in these proteins lead to loss of transparency and cataract remains uncertain. More insoluble protein is found in older than in younger lenses and this is further exacerbated in lenses with cataract [4]. The causal link between insoluble protein, protein aggregation and cataract formation may appear obvious but it should be noted that when analysis of proteins is conducted, this requires extraction of the proteins from the lens. It is possible that older proteins and those in lenses which have cataract are more vulnerable to insolubilisation during the extraction procedure and may not necessarily be aggregating in the living lens. Indeed, a study on the Blue Eye Trevally fish lens found that although the lens was transparent, the amount of insoluble protein extracted was very high [5].

The relationship between the crystallins and water and how this relationship alters with age and cataract in the living lens is also uncertain. A proportion of water in the lens is bound to protein and a proportion is unbound or free water; the latter has been shown to increase with age [3]. The causal relationship of free water with age may or may not be indicative of cataract formation. However, unlike protein aggregation, free water can be measured in the living eye using Magnetic Resonance Imaging (MRI). The technique has been erroneously promoted as a means to measure refractive index in the living eye but this is incorrect as refractive index depends on protein concentration and the amount of total water: bound and unbound [1].

The circulation of water and delivery of nutrients to the lens cells is thought to be dependent on a gradient of hydrostatic pressure involving aquaporin proteins [6]. Measurements of the developing zebrafish lenses indicate that the predominant aquaporin, aquaporin 0 plays a critical role on the formation of the refractive index gradient and hence is likely to contribute to the hydrostatic pressure gradient, maintaining a healthy physiological environment and consequently an undisturbed refractive index gradient and a transparent lens [7]. The precise nature of this structure/function relationship remains unknown.

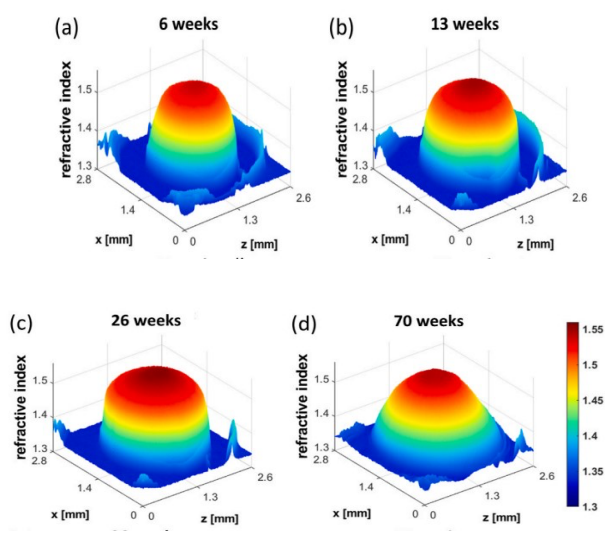

**Figure 1.** Three-dimensional refractive index plots for lenses of ARK mice from different ages : (a) 6 weeks, (b) 13 weeks, (c) 26 weeks, (d) 70 weeks. Modified from reference 13. K. Wang, Y. Pu, L. Chen, M. Hoshino, K. Uesugi, N. Yagi, X. Chen, Y. Usui, A. Hanashima, K. Hashimoto, S. Mohri, B.K. Pierscionek, "Optical Development in the Murine Eye Lens of Accelerated Senescence-Prone SAMP8 and Senescence-Resistant SAMR1 Strains", Exp. Eye Res. vol. 241, 109858, 2024

**Advances in Science and Technology to Meet Challenges**

The greatest advance in understanding the lens structure/function relationship that is needed is in imaging and the development of technologies that can allow viewing of the major structural entities: the lens proteins, crystallins and aquaporins in the living lens. Great strides will be made even if such technologies were available for measurements in the in vitro intact lens. This would address the major and fundamental unknown aspect of protein behaviour in the living, or at least in the intact lens. It would provide the answer to whether and how insoluble protein extracted from a lens is related to protein aggregation and/or other changes within the lens. This would lead to a far greater understanding of how the proteins and any changes to these proteins with age affect the refractive index gradient as well as what sorts of alterations to these structural entities may predispose to cataract.

Advances in optical measurements using synchrotron radiation and interferometric analysis have allowed accurate measurements of refractive index in very small developmental samples, both wild type and mutant [7-9] (Figure 1). This has led to leading research in the understanding of how the refractive index develops and changes with age in a range of species [7,10-12], which mutations affect the refractive index gradient and how this manifests as a disruption to the optics of the eye [7,9,13]. These measurements have also produced ground-breaking research in the testing of anti-cataract agents [14]. A translational of these capabilities into the living lens would greatly enhance advances in measurement of optics of the eye.

A further challenge lies in the understanding of how the aquaporin proteins fulfil their function of allowing water and nutrients to enter the inner lens and how this is maintained as the lens continues to grow. There may also be transient effects on water/nutrient transport when the lens changes shape during the process of accommodation. The importance of systemic nutrition and how this influences the health of the lens cannot be underestimated. The health of the lens is highly dependent on a balanced diet and there is a growing interest in certain foods, such as those rich in polyphenols, as potentially protecting the lens from opacification [15]. The relevance of diet to lens health and accurate monitoring of its effects on the lens presents both a challenge and an opportunity.

**Concluding Remarks**

The lens is essentially a small sac of proteins and water contained within a semi-elastic membrane. The only pathology that is found in the lens is cataract and physiologically, the only ageing manifestation is a loss of capacity to focus on near objects. Yet, though structurally relatively simple, the lens is a sophisticated optical element that, to date, has been impossible to replicate in intraocular implants and the pathological and physiological changes continue to occupy scientific thought and endeavour. The key to understanding how the lens alters with age and what leads to cataract is a far greater knowledge of the behaviour of the crystallin and aquaporin proteins and how this behaviour affects function. The vast amount of research conducted on these proteins has required them to be extracted from their environment in the lens. How these proteins interact with other protein classes and with water within the intact living lens will require innovations in imaging technologies that do not currently exist. To develop such technologies is the greatest challenge for lens researchers in the future.

## 6. Lens models of the human eye

Jesús E. Gómez-Correa and Sabino Chávez-Cerda

Instituto Nacional de Astrofísica, Óptica y Electrónica, Coordinación de Óptica, Tonantzintla Puebla 72840, Mexico

[jgomez@inaoep.mx ; sabino@inaoep.mx]

**Status**

A schematic eye is a mathematical or physical model that represents the fundamental optical properties of the real human eye. In essence, it is a simplified version of the eye, conceptualized as an optical system composed of two lenses: the cornea and the crystalline lens (or simply the lens) [1]. The lens is the most fascinating and complex optical element of the eye. This complexity stems from its nature as an asymmetric Gradient-Index (GRIN) lens, where the refractive index decreases from the core to the surface. A remarkable property of the lens is its ability to alter both its shape and its GRIN distribution during the accommodation process, enabling it to focus effectively on objects at varying distances onto the retina. Thus, the lens is also dynamic [2]. Furthermore, the lens's complexity is enhanced by its continuous growth throughout life, which causes its shape and GRIN distribution to change with age. In general terms, the lens is a dynamic and asymmetric GRIN lens that is age-dependent [3].

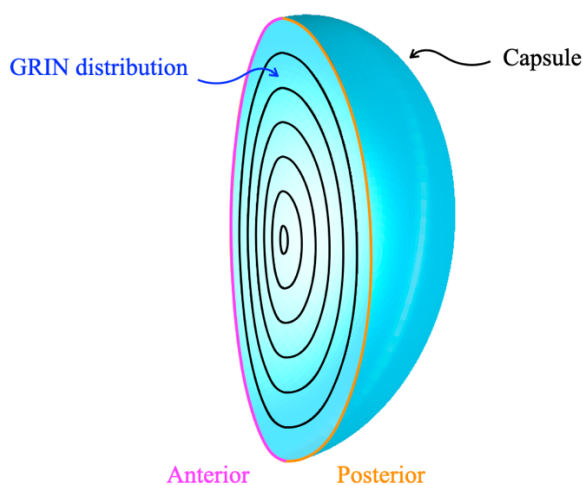

**Figure 1.** The anterior and posterior surfaces of the crystalline lens, along with its gradient refractive index (GRIN) distribution. The GRIN smoothly transitions from a higher refractive index in the central nucleus to a lower index toward the peripheral cortex. The anterior and posterior surfaces exhibit different curvatures, contributing to the lens's optical properties.

Over the last centuries, different lens models have been proposed and their accuracy and complexity have continued to evolve. Early lens models used simple spherical refractive surfaces with a constant refractive index [4-7]. Some more recent models still focus solely on surface geometry [8-10], but these surfaces are no longer spherical. Other models concentrate on representing the GRIN distribution [11,12], where the refractive index varies throughout the lens. Additionally, more robust models incorporate the accommodation process and/or age dependence [13-19]. These models can be mathematically represented by one or two functions due to the asymmetry of the lens. In the case of two functions, one represents the anterior part and the other the posterior part of the lens (see Figure 1) [14-17,19,20]. If a single function is used, it represents both parts [8,9,11-13,18]. For the GRIN distribution model with two functions, they must obey continuity conditions to ensure that the transition between them is smooth. The advantage of single-function models over others lies in the fact that they guarantee the smoothness of the refractive index profile of the lens throughout its entire volume without the need to impose boundary conditions.

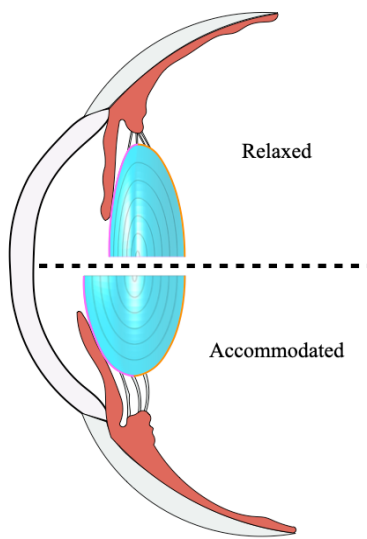

**Figure 2.** Representation of the accommodation process in the human eye. The relaxed state (top) shows the ciliary muscle at rest, causing the lens to flatten for distant vision. In the accommodated state (bottom), the ciliary muscle contracts, allowing the lens to thicken and increase its curvature for near vision.

The crucial function of the crystalline lens in fine focusing during the accommodation process is achieved through changes in its shape and GRIN distribution (see Figure 2) [21]. This ability is the result of an evolutionary process that has proven challenging to replicate in lens models. Most existing models of the crystalline lens treat its external shape independently of the refractive index and subsequently adjust imaging properties through optimization processes. One reason for this is that creating a model of the crystalline lens that simultaneously incorporates both parameters has been a formidable task. However, there are now two models that incorporate both parameters: the Poisson-Gauss lens model [13,22] and the AVOCADO model [19].

**The Poisson-Gauss lens model:** In 2020, Jaimes-Nájera et al. proposed a physiologically realistic GRIN model of the crystalline lens based on a single function that accurately describes different accommodative states [13,22]. This model provides the corresponding refractive index distribution and external shape of the lens capsule by adjusting a single parameter associated with the function of the ciliary body. Changes in equatorial and total axial lens thicknesses, as well as aberrations, were found to be within reported biometric data ranges.

**The AVOCADO model:** In 2016, Sheil and Goncharov introduced a new model that describes the age-dependent human lens using two independent axial and radial gradient index distributions [19]. Recognizing that the stability of the optimization process to match any proposed gradient index variation to experimental data directly depends on the degrees of freedom in the mathematical representation of the gradient index, the authors chose the lens's external shape as a basis to constrain the gradient index distribution during optimization, thereby minimizing the number of free parameters. They demonstrated that the spherical aberration, calculated through exact ray tracing, aligns with experimental data.

**Current and Future Challenges**

Due to all properties of the lens, generating a model that encompasses all its characteristics is very challenging [3]. The primary challenge lies in modelling both the external and internal structures of the lens. These models must include the accommodation process and age dependence, providing realistic predictions of lens power and aberrations. Additionally, all parameters and their optical properties must be consistent with experimental measurements.

One of the major current and future challenges is fully understanding the modification of the GRIN structure and the lens capsule during the processes of accommodation and aging [23]. Achieving a comprehensive understanding of the lens in these contexts would lead to a better grasp of how the optical properties of the entire eye change, given that the lens is the most complex optical structure involved in both processes. The lack of thorough knowledge of these processes has, over the years, led to numerous efforts to develop models that accurately reproduce and explain all the changes in the gradient index structure of the lens during accommodation and aging [24,25].

To obtain better models, more accurate in vivo measurements are necessary, which is challenging due to the lens's position behind the iris, making direct measurements of its shape and optical properties difficult. Additionally, the intrinsic movement of the human eye complicates error-free measurements.

Most lens models are developed using averaged measurements of parameters and their optical properties. However, this also presents a challenge since these averages are typically derived from a population in a specific region, which may not accurately represent other populations. This raises the need for personalized measurements, paving the way for customized lens models tailored to individual needs [3]. However, some parameters cannot be obtained from a single measurement, so we must rely on existing literature, which often reflects averages from certain populations.

**Advances in Science and Technology to Meet Challenges**

Advancements in lens modeling largely depend on the development of clinical tools and software that can accurately measure key parameters, such as the surface geometry and refractive index distribution within the lens. However, these tools must be capable of measuring these parameters during the process of accommodation. This capability would lead to more accurate lens models and enhance our ability to predict how the optical properties of the lens change with accommodation and over time.

For instance, a deeper understanding of accommodation in these models could lead to the design of more advanced intraocular lenses and other optical devices, such as contact lenses, that adapt more effectively to the human eye's needs throughout life. Additionally, the ability to accurately model lens aging could open new avenues for early intervention before visual issues like presbyopia or cataracts arise, ultimately improving the quality of life in the aging population.

Another significant advancement lies in the evolution of artificial intelligence (AI), which has the potential to revolutionize the creation of more precise lens models. AI, with its ability to analyze large volumes of data and detect complex patterns, can significantly enhance the accuracy of lens models, enabling more detailed and realistic simulations of how the GRIN structure and lens capsule change during accommodation and aging.

Moreover, AI could streamline the generation of custom intraocular lens designs more efficiently. By utilizing machine learning algorithms, it would be possible to create lenses tailored specifically to the visual needs of each individual, taking into account factors such as unique ocular geometry and variations in accommodation response. This would not only optimize postoperative outcomes but also allow for unprecedented precision in addressing visual problems, ultimately improving patients' quality of life.

**Concluding Remarks**

In general, a deeper understanding of the processes of accommodation and aging will allow the development of more accurate lens models. These improved models, in turn, will lead to advances in the design of intraocular lenses, which will translate into a better quality of life for people affected by lens-related conditions such as presbyopia and cataracts. With more precise models, we will be able to create more personalized and effective optical solutions, addressing visual problems more directly and efficiently.

**Acknowledgements**

This research was supported by Instituto Nacional de Astrofísica, Óptica y Electrónica (INAOE)

Advances in Visual and Physiological Optics Roadmap, *Journal of Optics*

## 7. Ray tracing in the human eye lens
Alexander V. Goncharov
School of Natural Sciences, University of Galway, Ireland

[ alexander.goncharov@universityofgalway.ie ]

**Status**

Ray tracing in the human eye lens has its beginnings in the early study of optics, with foundational work by Johannes Kepler in the 17th century, who first described the eye as an optical instrument. Kepler's insights into the role of the crystalline lens in focusing light on the retina influenced later scientists, including his contemporary René Descartes, and later Hermann von Helmholtz, who further developed the understanding of the eye's optics, providing a theoretical basis for the study of accommodation and the development of corrective lenses. Helmholtz's detailed exploration of light refraction through the eye, although not strictly ray tracing as we understand it today, paved the way for later advancements [1]. The emergence of computational ray tracing methods in the 20th century enabled increasingly accurate calculations of light paths through schematic models of the human eye, incorporating more realistic features such as aspheric surfaces and a gradient refractive index (GRIN) lens. These methods transformed the way scientists predict the path of light as it traverses the eye, particularly within the complex GRIN structure of the crystalline lens.

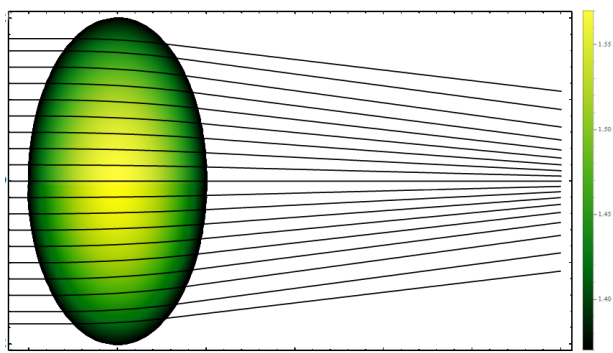

**Figure 1.** The curved ray paths in a quadratic GRIN lens, for which the refractive index at the lens surface is $n_0$=1.37 and at the centre $n_{max}$ =1.57, the large value $n_{max}$ is chose to emphasise the ray bending effect in the lens.

Nowadays, ray tracing in the human eye lens is an invaluable tool in both research and clinical practice [2]. While paraxial ray tracing predicts optical power, image size, and its position, exact ray tracing allows for the estimation of ocular aberrations and their effects on vision quality in eyes with real (finite) pupil sizes. In the latter case, the main challenge is performing ray tracing in the GRIN lens, where light rays travel along curved trajectories, as shown in Figure 1. Determining these trajectories usually requires numerical iterative methods [3,4].
Therefore, ray tracing is an essential for understanding the origin and impact of lens aberrations on both central and peripheral vision. It is also crucial for studying the effects of aging on lens shape and its internal GRIN profile (lens paradox) [5], the reduction in accommodation amplitude (presbyopia) [6], and overall image quality [2]. Ray tracing is also needed for developing customized methods of vision correction. These methods include the design of spectacle and contact lenses, intraocular lenses, and the refinement of refractive surgeries like ray tracing-guided LASIK. As optical models of the eye become more sophisticated and realistic, assisted by exact ray tracing, they will enable the development of even more precise corrective lenses and surgical techniques, potentially allowing for fully personalized vision correction solutions.

**Current and Future Challenges**

Modern ray tracing techniques combine geometric optics with wavefront analysis, enabling highly detailed simulations of how light propagates through the eye. In particular, the error in the ray path within the lens should not exceed a fraction of the wavelength of the light used. Exact ray tracing can be easily performed at real heights and large incidence angles in conventional optical systems consisting of lenses with a constant refractive index. However, for eye models with a GRIN lens, exact ray tracing becomes approximate, requiring many iterations to achieve the same level of accuracy as in conventional lenses. For some existing eye models with a GRIN lens, it is possible to analytically calculate the paraxial properties of the lens, such as its power and back focal distance [7, 8]. However, performing ray tracing at finite heights analytically is only feasible for simpler models featuring a quadratic GRIN profile [9]. For more advanced lens models, numerical methods are required to solve a non-paraxial differential ray equation [4,10].

Time-efficient finite ray tracing in GRIN media relies on various numerical methods, but these methods do not provide an analytical solution; at best, ray tracing in the reverse direction can yield the same residual error [11]. The absence of analytical methods prevents solving the inverse problem—namely, recovering the GRIN distribution of the lens given the incoming and outgoing light beams (or wavefronts) after they have passed through the lens. The challenge of reconstructing the lens using ocular tomography [12,13] involves dealing with noisy experimental data and non-unique solutions to the problem, which may lead to unrealistic GRIN profiles [14]. Using layer-by-layer finite ray tracing through a lens with numerous constant refractive index shells can introduce some realism regarding discontinuities that may occur in the GRIN lens profile [15]. However, a lens model with a large number of shells that closely mimics the crystalline lens with a smooth GRIN profile still does not provide a solution to the inverse problem of lens reconstruction due to the sheer number of discreet refractions involved [16]. Ray trajectories with continuous refraction in future GRIN lens models might be describable semi-analytically, achieving sufficient accuracy while also enabling the solution of the inverse problem. Alternatively, one could apply brute-force optimization methods to obtain a solution, though it would heavily depend on the constraints and the choice of optimization variables in the model [17].

**Advances in Science and Technology to Meet Challenges**

The development of more accurate patient-specific models of the eye requires advances in optical methods for reconstructing eye parameters, particularly those of the lens. One of the important roles of ray tracing is to accurately analyze the optical properties of a given lens. However, a more challenging application is inferring the GRIN structure of the lens from experimental biometric data. This task is difficult in vitro [18] but becomes significantly more challenging in situ when reconstructing the optical system of a living eye.

In this context, fast and accurate ray tracing through the lens is essential not only for predicting how the image is formed on the retina (a forward problem) but also for imaging the lens itself, particularly its posterior part, which becomes distorted by the unknown GRIN structure (an inverse problem). This distortion effect must be considered when imaging the lens using OCT or Scheimpflug slit systems. These imaging techniques are not accurate enough to measure the lens shape with fraction-of-a-wavelength accuracy; at best, they can achieve a resolution of a few microns. However, they can still be useful for constraining the external lens shape and the shape of its nucleus. Measuring ocular aberrations across the visual field with wavefront sensing or laser ray tracing [19] ultimately provides a more accurate method for reconstructing the optics of the eye and lays the foundation for wavefront tomography. Additionally, measuring the optical path length (time of flight) in the eye will provide further constraints on the axial optical thickness and the eye's overall length. An interesting alternative is to use OCT imaging at different off-axis angles [20].

The challenge of gathering biometric measurements in the living eye lies in the unavoidable changes that may occur if the measurements are done sequentially with some delay. Additionally, all measurements should be performed with instruments operating at the same wavelength; otherwise, chromatic effects might complicate the reconstruction process. Wavefront sensors operating across wide visual fields have already been used for retinal imaging systems assisted by adaptive optics.

However, integrating additional imaging modalities into a single instrument for wavefront sensing and simultaneously measurements that provide valuable constraints on the shape and position of the lens in the eye, remains to be done. Ultimately, this will enable researchers to capture more detailed information about the shape and optical characteristics of the crystalline lens, which in turn will allow for the development of more subject-specific eye models and potentially more efficient methods of ray tracing through a GRIN lens.

**Concluding Remarks**

The most complex optical element in the eye is the crystalline lens, which features a gradient index (GRIN) structure that plays a crucial role in image formation on the retina. To analyze the imaging properties of the GRIN lens, numerical methods for ray tracing through the lens are employed. However, due to the lack of an analytical solution for ray trajectories in the GRIN medium, the inverse problem of reconstructing the lens in situ with current state-of-the-art imaging modalities remains unsolved, as the solution heavily depends on the GRIN lens model used. For future, more realistic patient-specific eye models—requiring fraction-of-a-wavelength accuracy for personalized vision correction—an integrated approach combining wavefront tomography and OCT imaging, assisted by efficient exact ray tracing, will be needed. If successful, this will enable the development of more customizable corrective lenses and surgical techniques for fully personalized vision correction.

**Acknowledgements**

Open access funding provided by Irish Research eLibrary. This research was also supported by the Hardiman Research Scholarship at the University of Galway, Ireland.

## 8. Photoreceptors of the human eye

Amy Fitzpatrick and Brian Vohnsen
Optics Group, School of Physics, University College Dublin, Ireland

[Amy.Fitzpatrick1@ucdconnect.ie ; Brian.Vohnsen@ucd.ie]

**Status**

The colour sensitivity of the human eye in photopic conditions originates in the 3 cone types of the retina that are most densely packed at the fovea. Here, the green and red sensitive M- and L-cones dominate with just a small fraction being blue sensitive S-cones (absent from the fovea centralis). This combination results in a sensitivity peak at 555 nm wavelength in daylight conditions. In turn, the rod photoreceptors, that have their peak density along a ring at an approximately 20° eccentricity, provide limited vision in scotopic conditions [1]. The visual pigments are packed in membrane infoldings in the outer segment of the cones and in stacked discs of the rods. Pigments renew at an approximately 2-week interval, with the oldest discs and pigments being shedded and engulfed by the dark retinal pigment epithelium (RPE) cells located just beyond the photoreceptors in the process of phagocytosis [2]. The photoreceptors absorb only a fraction of the incident light with the remainder largely absorbed by the RPE layer. Backscattering of light is very low, which is why the pupil appears dark. Yet, advanced fundus imaging techniques, often with adaptive optics and infrared light, allow direct visualization of the cones and rods in the living eye.

Photoreceptors have long been considered to guide light due to a strong dependence on the angle of incidence for light at the retina [3,4]. Nevertheless, there is evidence that the directionality may rather relate to the pointing and arrangement of the elongated cone cells in combination with a focusing effect by mitochondria in the ellipsoid [5,6]. This explains the angular dependence of the Stiles-Crawford effect of the 1$^{st}$ kind [7-9] as well as the perceived change in hue for oblique light known as the Stiles-Crawford effect of the 2$^{nd}$ kind [10,11] without enforcing waveguiding. In this interpretation, both are consequences of leakage of light from the individual outer segments with potential crosstalk between adjacent cone photoreceptors. It also explains the lack of a Stiles-Crawford effect in scotopic conditions, as leakage of light from one rod, can be captured by opsins in a neighbouring rod thereby flattening the angular response.

The detailed understanding of photoreceptors as an integral part of the optics of the eye may explain why each eye is best adapted to its own pattern of aberrations [12]. This opens ways for personalized optical corrections without prolonged adaptation periods. The temporal dynamics of photoreceptors at both short [13-15] and long timescales [16-18] are of significant interest. Photoreceptors react to light exposure changing the optical path length as explored with optoretinograms [13,14] that may potentially be used as biomarker for disease and retinal change [15]. Fast ocular motion, drift, tremor and saccades play a direct role for our visual system at a sub-photoreceptor level [19,20]. Further advancement in the field of photoreceptor optics and phototropism may hold the key to understand adaptation to aberrations at the retinal level [9,21-23] prior to neural responses with vital new inputs into emmetropization and myopia prevention [18,24,25].

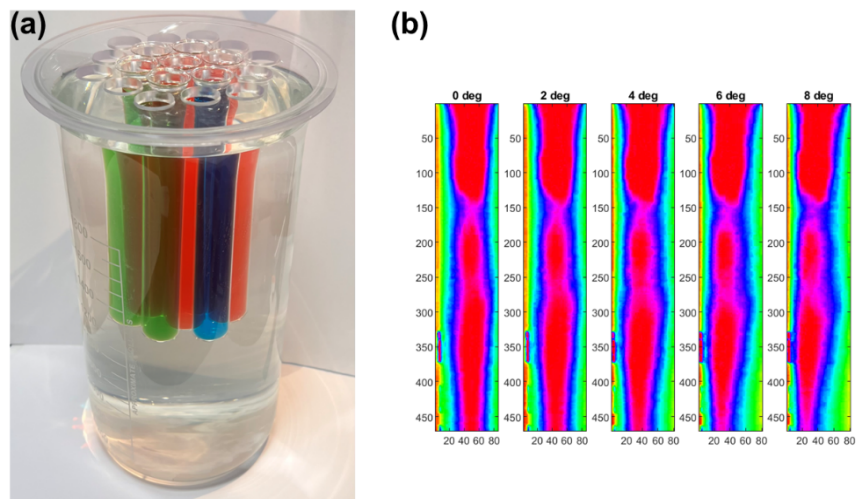

**Figure 1.** Macroscopic model of the foveal photoreceptor mosaic (a) with test tubes and food dyes to emulate the S, M and L cones of the human retina. The model is being used to (b) test the hypothesis of the Stiles-Crawford effects being mostly due to leakage of light.

**Current and Future Challenges**

We have detailed knowledge about the cone and rod photoreceptors in the adult eye, and the impact of ageing and retinal disease, but our knowledge about retinal changes during childhood remains limited [25,26]. Our knowledge about changes occurring to the photoreceptors over short [13-15] and long timescales is still also sparce [16-18]. From optical coherence tomography, we know that photoreceptors respond to absorption of light by minute changes to the optical path (the integrated combination of length changes and refractive index alterations). This originates in the photoisomerism triggered by light absorption followed by slower osmotic swelling, but there are also small temporary photothermal changes caused by heating of the outer segment and the RPE cells. At a slower time scale, photoreceptors show a certain phototropism whereby they adapt their orientation over the course of approximately 2 weeks [16,17]. The renewal of pigment layers in the outer segments will likely happen in such a way that cells maximize their light capture. Each visual pigment does not perceive an image, but only triggers by the detection of photons with the most efficient configuration being that outer segments are pointed towards a common pupil point [27].

In the process of emmetropization and ocular growth in children, the cones and rods across the retina reshape and become denser likely linked to the optics of the anterior eye. As the absorption of light in each outer segment is limited to a fraction of the incident light, one may ask "why are outer segments not longer to capture more of the light?" It is likely an optimization process of photon capture and visual acuity. Longer outer segments would secure higher photon capture but would also increase the likelihood of light leakage and crosstalk degrading vision. Ultimately, it would cause challenges for the cellular energy supply and the process of phagocytosis. It appears plausible that adaptation of the still developing photoreceptors in infants and children would be best matched to outdoor lighting conditions where the pupil is smallest helping to prevent myopia onset [18]. Evolution has prepared our eyes for outdoor life. Here, the pupil is small, and the risk of myopia onset is smaller. Clearly, photoreceptors should be understood as an integral part of the eye on par with the refractive properties of the anterior eye. If such knowledge can be used to find new methods to limit myopia onset and progression remains to be seen [24,26]. Such knowledge may also help to understand how the eye adapts to refractive changes and thereby provide a better fit for subjective corrections from the moment new contact lenses or spectacle glasses are being used for the first time.

**Advances in Science and Technology to Meet Challenges**

The past decades have witnessed impressive improvements in technological capabilities to image the retina and photoreceptors of the living eye. The development of optical coherence tomography to visualize the individual layers across the retina and the use of adaptive optics to provide diffraction-limited transverse resolution at a cellular level have been vital ingredients for these achievements. We would still need more advanced imaging technology to visualize the internal structure of the cone and rod photoreceptors with higher precision, to learn where changes happen during ocular growth and where new refractive designs make a direct impact on the structure and function of the photoreceptors, for example, in myopia control. The development of high-resolution optoretinograms will likely be a key component to achieve this and will help us distinguish between true optical changes at the retina from that of neural origin.

In some of our work we use retinal simulators [28] from the microscopic to the macroscopic scale. One such example is shown in Figure 1 where an array of test tubes with dyes are used to simulate the Stiles-Crawford effects with an effective directionality determined from the scattering of light incident at different angles based on the antenna model of the photoreceptors [5]. We are also conducting novel research of light absorption in the photoreceptors using both ray optics and electromagnetic modelling as shown in Figure 2. While we still need more details for the developing eye and retina in children, we have now at our disposal advanced modelling techniques that can be used to test hypotheses and optical corrections in numerical ocular and retina models, as well as with physical models of the eye and retina.

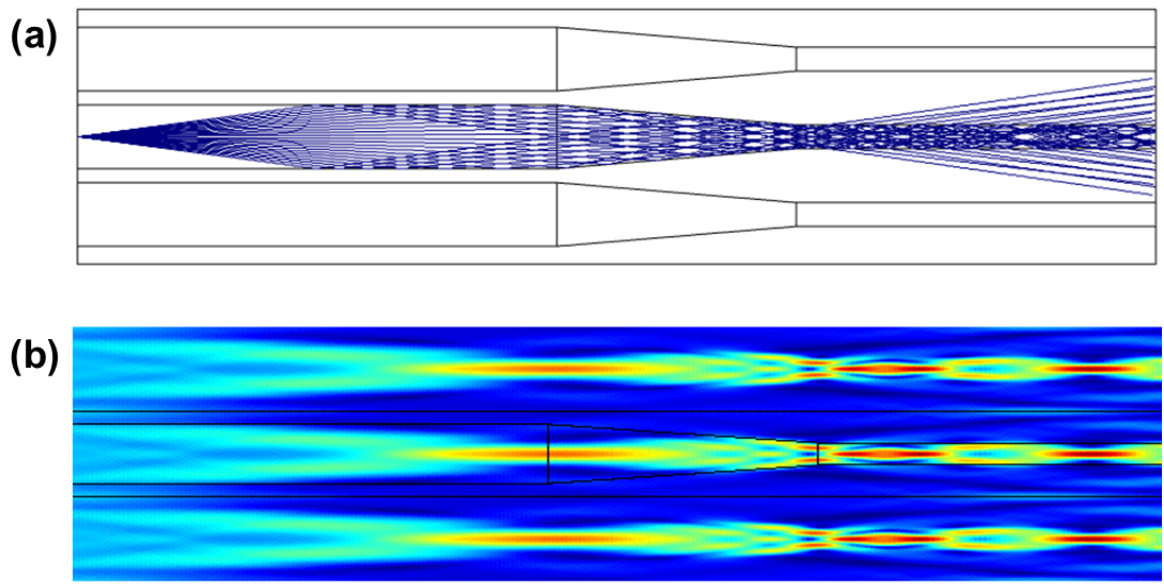

**Figure 2.** Electromagnetic calculations of light leakage from 3 adjacent peripheral cones using (a) ray optics and (b) wave optics, respectively. Only the ideal case of on-axis incidence is shown, but the model is equally capable of analysing oblique light incidence.

**Concluding Remarks**

The journey towards understanding the optics of photoreceptors has been long and complex. Although a directionality in the light transmission was recognized early on, it was the discovery of the Stiles-Crawford effect that triggered a range of experiments aimed at understanding how vision responds when light is obliquely incident onto the retina. The focus was on Maxwellian illumination which may have delayed the discovery of a substantially stronger apodization by the integrated Stiles-Crawford effect in normal viewing conditions [9,21,22]. As we understand the optics of the retina better, it has become clear that not only does the retina and visual pigments adapt to changes in illumination at both fast and slow time scales [13-18] but equally it plays an important role for the overall ocular optics in line with the refractive optics of the anterior eye. With this knowledge, we expect that future ophthalmic products may use this to support improved and personalized vision aids, and ultimately lead to new diagnostic tools for earlier detection of vision loss. New retinal implants to restore vision for the blind could utilize this by emulating the optics of the photoreceptors.

**Acknowledgements**

This research has been realized with financial support from the UCD SIRAT scheme and H2020 ITN MyFUN grant agreement No. 675137.

## 9. Retinal imaging: Where we are and where we are going

Jessica I. W. Morgan[1,2]

[1]Scheie Eye Institute, Department of Ophthalmology, University of Pennsylvania, Philadelphia, PA, USA

[2]Center for Advanced Retinal and Ocular Therapeutics, University of Pennsylvania, Philadelphia, PA, USA

[ jwmorgan@pennmedicine.upenn.edu ]

**Status**

Retinal imaging has long been an important part of the ophthalmic exam as it allows visualization and documentation of retinal phenotypes that aid in the diagnosis, prognosis and treatment of ophthalmic diseases including age-related macular degeneration, glaucoma, diabetic retinopathy, and inherited retinal dystrophies [1]. Imaging in the eye has also served as a technique to visualize disease in the vascular and neurological systems, as changes in the retina have been observed in conditions of hypertension, diabetes, dementia and Alzheimer's Disease among other systemic and neurological diseases [2-4].

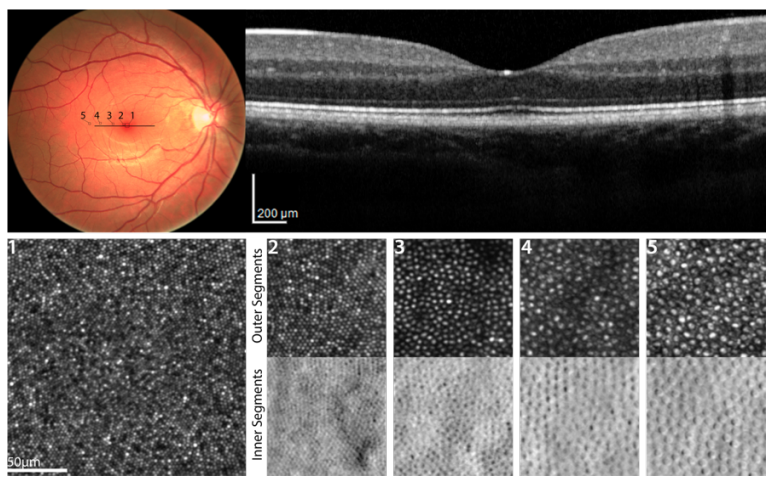

**Figure 1.** Color fundus photograph (top left) of the normal macula, cross-sectional OCT b-scan through the fovea (top right) at the location marked by the black line in the fundus photograph, and adaptive optics images of the photoreceptor inner segment and outer segment waveguided mosaics (bottom) at the numbered locations marked in the fundus photography.

Numerous technologies abound for retinal imaging, many of which are now considered standard care in ophthalmic practice, others of which are considered advanced tools used only in specialized cases or research settings. The following paragraph provides a general overview of the most commonly used techniques for retinal imaging, but broadly speaking, any imaging technologies employed in microscopy can been applied to retinal imaging, where the natural optics of the eye serves as the objective lens of the microscope. The historical gold standard is fundus flash photography which provides a color view of the macula, optic disk, and retinal vasculature5 (Figure 1, top left). Scanning laser ophthalmoscopy (or line scan ophthalmoscopy) employs a light source that is focused on the retina which is then physically moved across the retina to enable en face video-rate imaging of the scanned area [6]. An advantage of scanning laser ophthalmoscopy is that the light returning from the retina can be collected through a confocal aperture thereby reducing scattering from the ocular media. A disadvantage however, is that any eye motion that occurs during the scanning acquisition will cause distortions within the obtained retinal image. More recently, optical coherence tomography (OCT) is a technique arising from interferometry that provides cross-sectional volumetric imaging of the retina [7] (Figure 1, top right). A major advantage of OCT is its high resolution in the axial direction, which enables visualization of the retinal lamination. Methods for enhancing the contrast of retinal features, such as fluorescence or motion contrast, are also available [5]. Fluorescence techniques in combination with fundus photography or scanning ophthalmoscopy enable visualization of intrinsic retinal fluorophores such as lipofuscin and melanin in the retinal pigment epithelium or extrinsic fluorophores delivered to the eye via the vasculature such as fluorescein and indocyanine green. Motion contrast methods, such as OCT-angiography involve the comparison of multiple sequentially acquired images, where any differences between consecutive images are attributed to blood flow. Each of these techniques provide complementary visualizations of the retinal phenotype and inform clinicians about the health of the retina under examination.

Retinal imaging through the natural optics of the eye however, is limited by both the size of the pupil and the quality of the eye's optics. Diffraction theory would suggest that as the pupil gets larger the resolution of the system would improve. However, as the size of the pupil increases, so too do the eye's optical aberrations. Adaptive optics (AO) is a tool for first measuring and then compensating for the eye's optical aberrations [8]. This technique enables diffraction limited resolution even when the eye's pupil is fully dilated. AO has been combined with fundus photography [8], scanning laser/light ophthalmoscopy [9], OCT [10], fluorescence imaging [11], and motion contrast [12] among other imaging technologies. With these techniques, investigators have been able to visualize individual cells within the living retina including the cones and rods within the photoreceptor mosaic (Figure 1, bottom), the retinal pigment epithelium (RPE), ganglion cells, individual white and red blood cells, and macrophages among others [1].

The vast majority of retinal imaging is used to inspect the structure of the retina in health or disease. Recent techniques however have enabled investigations into the function of observed structures. Like structural imaging there are multiple complementary techniques for investigating function, including methods which incorporate subjective responses of perception such as perimetry or acuity [13], or physiological measures such as blood flow. A new, rapidly emerging field includes optoretinography, which involves measurement of an optical signal originating in the retina in response to a visible stimulus. Most studies in this field to date have investigated the optoretinogram (ORG) of photoreceptors [14,15], though similar experiments have targeted measurements of bipolar and ganglion cells [16]. Current theory suggests the phase of light reflected from cell surfaces/interfaces is altered following light exposure by both contraction and expansion of the cell's membrane and that these changes correlate with mechanisms involved with phototransduction in the photoreceptors or activation of inner retinal neurons [17,18]. Though it is still a new technique, optoretinography has demonstrated strong potential for high resolution, noninvasive, objective measurements of physiological cell function in the living retina [19,20].

Clinical trials investigating experimental therapeutic approaches usually are required to demonstrate a functional benefit for the trial to be considered successful. In some cases, structural results from retinal imaging may be used as the trial's primary outcome measure, in particular if the structural phenotype being used is known to correlate with better vision. Examples of structural phenotypes that correlate with vision include the extent of the ellipsoid zone (one of the retinal layers observed on OCT attributed to photoreceptors) or the extent of geographic atrophy (visualized as a loss of autofluorescence attributed with loss of the RPE). Studies which strive to understand structure-function relationships therefore are crucial for interpreting the disease phenotypes observed on images and will continue to be important for identifying structural measurements that are appropriate for use as outcome measures in clinical trials.

**Current and future challenges**

Retinal imaging is playing an increasingly important role in clinical care, yet there are still numerous challenges to overcome. For one thing, imaging in elderly patients or patients with disease usually is more difficult than imaging in young, healthy controls. Patients many times have abnormal eye movements that can reduce the success rate of imaging and lengthen the time spent acquiring images. Further, patients may have abnormal optics, cloudy media, or cataracts which hinder imaging. In addition, the structures under investigation may show abnormalities which may make the interpretation of findings more difficult. This is further complicated by the first date of presentation for many individuals. Indeed, many times irreversible vision loss has already occurred when a patient presents for initial imaging. This begs the question of whether retinal imaging should become a part of standard preventative medical care and if so, using what modalities and at what age(s)?

The time spent on data acquisition and analysis is also a major limiting factor for multimodal retinal imaging. High resolution multi-modal imaging typically requires pupil dilation and access to multiple different imaging devices, resulting in extended appointments for patients. In addition, quantitative analysis is a burdensome task as reading or annotating images requires significant time by trained individuals and therefore most analysis takes place in a post-processing environment. These problems highlight the need for validated, automated algorithms capable of extracting relevant biomarkers in real time and ideally early in disease processes when intervention is more likely to succeed. Collaborative teams including engineers, scientists, and clinicians will be needed to identify the most successful techniques and approaches for these problems.

**Advances in science and technology to meet challenges**

With sufficient attention and resources solutions to these issues will be found. Already, normative databases of multi-modal retinal images are becoming available. Large databases of structural and functional imaging results that take into consideration variations in age, sex, and ethnicity will be needed to affirm differences between control and disease features and provide testing grounds for new algorithms, including those developed using machine learning and artificial intelligence approaches, to identify biomarkers of disease. To be most impactful, future studies should take an active approach to data sharing by providing access to images for broad use and algorithm development.

Improvements in image acquisition, such as eye/motion tracking and real-time registration also are under development to enable better, faster acquisition in challenging patient populations. These technical improvements will also aid in expediting studies to develop functional applications of retinal imaging such as optoretingraphy.

**Concluding Remarks**

As a practical matter, the structure and economics of clinical trials and medical care puts a high value on biomarkers that reveal positive effect or lack thereof reasonably soon after application of therapy.

Such biomarkers are best formulated in the context of the most precise possible characterizations of disease progression. To this end, retinal imaging has provided quantitative tools for objective, noninvasive assessment of retinal structure and function in both health and disease. Thus, high resolution retinal imaging shows high potential to impact our understanding of disease mechanisms, progression, and treatment efficacy. Its importance in ophthalmic and systemic medical care will only continue to grow in years to come.

**Acknowledgements**

Funding provided by: National Institutes of Health (NIH R01EY028601, P30EY001583), F. M. Kirby Foundation, Research to Prevent Blindness.

``

## 10. State of the art of adaptive optics for evaluating retinal structure and function and healthy and diseased eyes

Austin Roorda

Herbert Wertheim School of Optometry and Vision Science, University of California, Berkeley, Berkeley CA 94720-2020

[ aroorda@berkeley.edu ]

**Status**

The saying "*there is no substitute for good optics*" applies to many fields - astronomy, photography, microscopy, and refractive corrections - but the field of ophthalmoscopy may make the strongest case. The eye optics is generally well-adapted for normal human vision, but ocular imperfections prevent imaging and optical interrogation of the visual system on a cellular level. Adaptive optics (AO) is a set of tools to measure and correct these imperfections and overcome these limits. AO technologies have been used effectively in ophthalmoscopes since 1996 [1] and have enabled dozens of revelations and discoveries in healthy and diseased eyes [2-4]. AO has proven to be an effective technology for many ophthalmic imaging modalities including full-field [1], scanning light imaging [5] and optical coherence tomography [6].

AO is useful because it enables near-diffraction-limited focusing through the eye's largest pupil size (highest numerical aperture) both on the way in and out of the eye. For large pupils, the diffraction-limited focused spot is on the order of single retinal cells and this level of access has driven a paradigm shift in how ophthalmoscopes are used to study vision in healthy and diseased eyes, effectively turning an ophthalmoscope into a microscope. And, like a microscope, there seems to be no limit in the arsenal of imaging methods and modalities that the use of AO will complement. Through the use of multiple AO-assisted techniques, nearly all retinal cell classes have been imaged *in vivo*, mostly in humans but also in animal models.

Laws of optical reversibility also mean that an AO correction is equally effective for light entering and exiting the eye. So, in addition to correcting light coming out of the eye to imaging structure, AO can be used to correct light going into the eye to test vision with diffraction-limited optics or with specific optical manipulations (e.g. adding aberration for increased depth of focus)[7] or to measure function on a cellular scale [8]. Importantly, systems that combine imaging and vision testing can be used to make direct relations between structure and function on that scale [9] in both healthy [8] and diseased eyes [10]. Functional testing falls into two classes, subjective and objective. Subjective tests include forms of microperimetry and visual acuity testing. Objective tests take multiple forms. Noninvasive approaches include the emerging field of optoretinography (ORG) for which the optical coherence-based techniques are the most advanced, providing an entirely new way to classify cone types [11] and see individual cone dysfunction in eye disease [12].

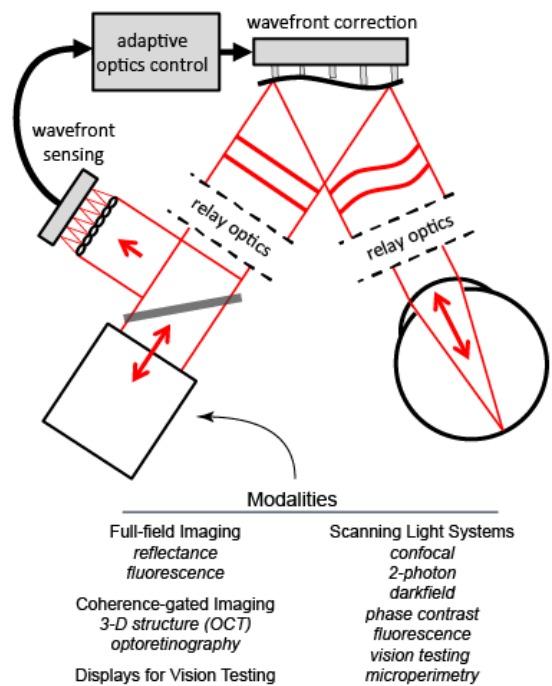

**Figure 1.** Generic schematic of an AO system for the eye with a list of all the known imaging and vision testing modalities that will benefit.

**Current and Future Challenges**

So how do we benefit from a microscope for the human eye? Initial discoveries enabled by AO were primarily structural: cone density in the fovea [13, 14], the packing arrangement of the three cone types [2], direct measures of cone structure for specific mutations [15], and microvascular structure [16]. But the biggest benefits will come from instruments that elucidate the retinal *function* in healthy and diseased eyes. A range of functional testing has already been done: measurement of the products of the visual cycle [17], blood flow in response to light stimulation [18], photoreceptor changes in response to light stimulation (ORG) [11], immune cell responses to retinal injury [19] and ganglion cell functional classification using optogenetics [20]. It is in this area that AO will make its largest impact. This is especially true in light of new treatments for eye diseases that are available today and are on the horizon. Having a tool that can evaluate retinal status prior to gene therapy, or the integration of stem cells in the host retina, or the functional benefits of treatments for retinal degeneration are essential. To do this accurately and quickly will help to accelerate treatments for eye disease. An even greater impact might be in early detection and diagnosis of disease, although that is a more difficult and daunting task than originally thought.

In the continued effort to relate structure to function in healthy and diseased eyes AO will continue to have a crucial role, even as the imaging modalities themselves will continue to evolve. For example, the notion that optical coherence-based methods could be used to reliably measure nm-scale change in cone outer segment length to light stimulation was largely unanticipated in the early days of AO and optical coherence tomography (OCT). The incorporation of AO into OCT enabled the classification the three cone types with more accuracy and in a shorter time than any previous approaches [11]. It is interesting to think of what new approaches are beyond the horizon.

**Advances in Science and Technology to Meet Challenges**

We are approaching 30 years since AO was first demonstrated for the eye, but it often feels like the technology is still in its infancy. Most AO systems are still custom-built by researchers and few companies have taken on the task of commercializing AO instruments and making systems more mainstream. The resolution benefits of correcting aberrations are unquestionable, so why is it not widely used? There are several reasons: First, the highest performance AO systems remain cumbersome, expensive and difficult to operate in comparison with other ophthalmic instrumentation. Second, the biggest benefits are realized for the largest pupil sizes (highest numerical aperture), but large pupils are also the most challenging to correct with AO. Third, the biggest benefits and most robust operation are for clear media. Collectively, these limits affect the ability to use AO to image the diseased or the ageing population who may have greater fixational eye movements, nystagmus, smaller pupils, more turbid media or intraocular lenses. In other words, the population that benefits the most are the most difficult to study. Finally, AO offers a microscopic view, which is an unprecedented scale for an ophthalmoscope. The scientific benefits are clear, but the immediate clinical usefulness is not. Until there is something learned from the microscopic view of a patient's retina that will inform a clinician on the course of treatment, the widespread need to have an AO device may always be low. This is further complicated by the fact that a consequence of the microscopic view is that the image field sizes are often small, in the opposite direction of many efforts in ophthalmoscopy to image an ever-wider field.

Despite these challenges, there is much scope for advances in technology that will make AO systems easier to use and more robust for a wide population. As those developments happen, the slope of the learning curve will steepen. With massive amounts of data on normal and diseased eyes, the ability to diagnose and track progression of a disease will continue to accelerate.

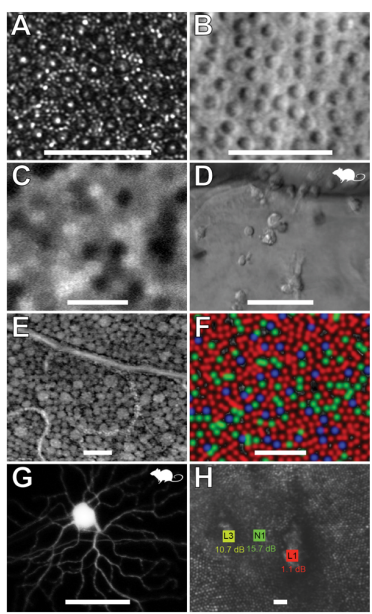

**Figure 2.** Selected images from AO systems. (A) AOSLO image of rods and cones using annular illumination. Adapted from [21]. (B) AOSLO phase contrast (split detector) image of the cone mosaic. Adapted from [22]. (C) AOSLO fluorescence imaging of the RPE mosaic using indocyanine green fluorescent dye. Adapted from [23]. (D) AOSLO phase contrast images of immune cells in a mouse retina. Adapted from [19]. (E) AO-OCT images of retinal ganglion cells. Adapted from [3]. (F) False-color image of the human trichromatic cone mosaic determined using optoretinography in a phase-resolved AO-OCT system. Adapted from [11]. (G) AOSLO fluorescence image of a mouse retinal ganglion cells. Adapted from [24]. (H) AOSLO microperimetry showing retinal function inside and outside lesions resulting from an accidental laser exposure. Higher dB scores indicate greater sensitivity. Adapted from [25]. Scale bars in every panel is approximately 50 micrometers.

**Concluding Remarks**

The advantages of *in vivo* microscopy are clear and AO in its current form or whatever future form it may take will continue to be the best technology to achieve that.

Over the past 30 years, those in the field have learned how to leverage AO in multiple modalities to image most cell classes in the retina, resolve dynamic functional processes, and explore vision on a cellular scale. Many discoveries have already been made, but the field is poised to make many more in healthy and diseased eyes.

**Acknowledgements**

Support from NIH grant: R01EY023591

## 11. Schematic models of the human eye

Jos J. Rozema[1,2] and David A. Atchison[3]

[1]Visual Optics Lab Antwerp (VOLANTIS), Faculty of Medicine and Health Sciences, Antwerp University, Wilrijk, Belgium

[2]Dept. of Ophthalmology, Antwerp University Hospital, Edegem, Belgium

[3]Centre for Vision and Eye Research, Queensland University of Technology, Kelvin Grove, Australia

[ jos.rozema@uantwerp.be; d.atchison@qut.edu.au ]

**Status**

Optical models of the human eye have been used to provide a framework for explaining optical phenomena in vision, predicting how refraction and aberrations are affected by change in ocular biometry and exploring optical limitations to vision. One category is the "toy train" type, a working tool that mimics the behaviour of real eyes but does not always attempt to be anatomically or mechanistically accurate. Schematic eyes fit into this type and can be described as mathematical entities that provide analytical descriptions of the eye's optical behaviour. Applications of schematic eyes include [1]: retinal image size calculations; predictions of refractive errors arising from variations or changes in eye dimensions; intraocular lenses power following cataract surgery; aberrations and retinal image quality with or without optical or surgical intervention; designing spectacles, contact lenses, intraocular lenses and corneal refractive surgery; and can be incorporated into the design of imaging instruments.

The simplest models are **reduced eyes** with a single refracting surface at the front of the eye. These are followed by **simplified eyes** with three refracting surfaces, one for the cornea and two for the lens, and then by four refracting surface models with two corneal and two lenticular surfaces. Next, allowance can be made for the gradient nature of the lens, first by nested shells with increasing refractive index towards the centre of the lens and then by true gradient index modelling. Distinction can be made between paraxial models and finite models. The former give accurate predictions of retinal image quality only for small pupils and for objects close to the optical axis. The latter are designed to improve predictions of optical quality both on and off-axis with the incorporation of media chromatic dispersions, by surface toricities, asphericities, tilts and decentrations, and by a curved retina.

Most schematic eyes to date represent population averages, with some developed for different situations or stratified by age, gender, ethnicity, refractive error, and accommodation. With the increasing availability of clinical instruments to measure ocular biometry, there is a move towards customization [2] that will be beneficial for refractive surgery applications and predicting the development of myopia at an individual level. Furthermore, we envisage a move towards more holistic models of wider appeal, involving not only optics but mechanical and other aspects of the eye and vision.

**Current and Future Challenges**

Although a wide variety of eye models has been published in the literature, most correspond to idealised averages of the biometry found in certain (sub)populations. Given the large biometric variations between individuals, such a *one size fits all* approach is clearly insufficient to produce a description that is inclusive for all eyes. Whether this is problematic depends entirely on the problem that the model is supposed to address. For the development of new optical solutions for, for example, myopia control, interindividual biometry variations are likely unimportant as the correction addresses the refractive error, which can be measured and corrected with great accuracy. When used in optical calculations, on the other hand, this issue may lead to greater problems as inaccuracies will compound and increase in magnitude the more the biometry of an eye deviates from those of the model being considered. A clear example of this is interocular lens (IOL) power calculation that is known to be inaccurate for eyes that are either too long or too short [3]. While this has prompted the development of new IOL calculation formulas dedicated to such cases [4], one may argue that more accurate eye models can lead to IOL formulas that are robust to any biometric combination.

Concerns such as these led to the development of **customised models** which integrate clinically measured data of a patient into a predefined structure, creating a hybrid model that is considerably closer to the original eye. But as not all parameters can be measured clinically and are taken from the generic predefined structure instead, some correlations between biometric parameters may be lost. This leaves room for further improvement.

An alternative approach is **generative eye models**, algorithms able to produce an unlimited amount of random, but plausible biometric data with the interindividual biometric variations found in the general population. Examples are the widely used wavefront model by Thibos et al. [5] and the more recent SyntEyes models [6,7]. Although these are currently not fully customizable to a patient's biometry, the output of generative modelling can be used in batch processing to e.g. assess the performance of new optical solutions in a wide biometric range.

Despite these advances, there are still important limitations to the realism of current descriptions as most models are purely optical and do not include other influences, such as mechanical pressure reshaping the eye globe or crystalline lens.

**Advances in Science and Technology to Meet Challenges**

The most important technological aspect that that can improve the accuracy of existing eye models lies in the development of clinical equipment and software to determine the ocular biometry more accurately, as well as new devices to measure currently difficult to obtain parameters such as the surface shape and gradient index of the crystalline lens, the shape of the retina and the refractive indices within the eye. Other potential improvements lie in reliable methods to measure the accommodative response, taken concurrently with biometry measurements, allowing more accurate accommodative models [8].

Besides improvements in the input data, there are many conceptual improvements to be considered as well, such as the influence of the **mechanical properties** on the ocular structures. As the eye is a physical structure, it is susceptible to stress caused by internal and external forces that lead to geometric deformations. But while there are many papers on mechanical eye models in the literature, these often do not consider the optical aspects, much as most optical models do not consider the biomechanical aspects. Combining both more often would be ideal, but this is hindered by the high computational cost of the finite element analysis software needed for the biomechanical calculations.

As developers are working to reduce these costs using artificial intelligence, this issue should eventually disappear.

Another advance is the development of models that describe the **growth and refractive development** of the eye [9], starting from an initial biometry state, such as at birth, and letting the eye grow virtually to adulthood under various behavioural and external influences. While interesting to study e.g. myopia development, current forms of such models contain only a minimal number of descriptive parameters and are based on literature until the 1990s. Further development by including more biometric parameters and the response of the various retinal layers and the choroid [10] can benefit the conception of new methods for myopia control.

Once such methods and models become available, clinical biometry data can be combined into an opto-biomechanical **virtual twin** of a patient's eye that can be used to try out refractive or surgical solutions virtually before performing them for real. Alternatively, one could perform **virtual clinical trials** of new techniques using either real or generated biometry data, which could form an intermediate step between animal testing and testing in the first group of real patients, thus providing an extra opportunity to identify issues beforehand.

**Concluding Remarks**

Beyond the current schematic eyes, which are mostly restricted to optics, a worthwhile goal is to develop a holistic (all-in-one) model that combines optics, mechanics and neural aspects. Some such models could also consider eye growth under various behavioural and environmental conditions. While complex, this could help address issues that involve many disciplines, such as to gain better understanding about accommodation or refractive development. Clinically, the development of virtual twins could help further personalise eye care, while generative models and virtual trials could become essential tools in the development of new surgical and refractive solutions.

Notwithstanding the above, it is generally best not to overcomplicate matters, but to critically consider the parameters and features to be included as often the simplest model to address a particular issue is best.

## 12. Modelling of ocular surfaces using different sets of functions

Juan P. Trevino[1,2] and Alejandra Consejo[3]
[1]Universidad Politécnica de Puebla
[2]Tecnológico de Monterrey, México
[3]Aragon Institute for Engineering Research (I3A), University of Zaragoza, Zaragoza, Spain

[ juan.trevino306@uppuebla.edu.mx, jptrevino@tec.mx, alejandra.consejo@unizar.es ]

**Status**

The interfaces among the optical elements of the eye predominantly determine the optics of the eye, so modelling ocular surfaces has been a central topic in ophthalmic optics. Efforts can be traced back beyond Gullstrand and LeGrand [1] who modelled the eye mainly considering spherical surfaces and the refraction indices within the eye. We may consider the most relevant surfaces to be the outer and inner surfaces of the cornea and crystalline lens. Although the latter is a graded index element, it has been often considered as a multilayered element [2,3]. In Figure 1 the simplest surface model can be observed. Other relevant surfaces that are not so obvious, such as the wave aberration function, the Point Spread Function, and the Modulation Transfer Function, are also important as they play a fundamental role in diagnosis and treatment for several eye conditions.

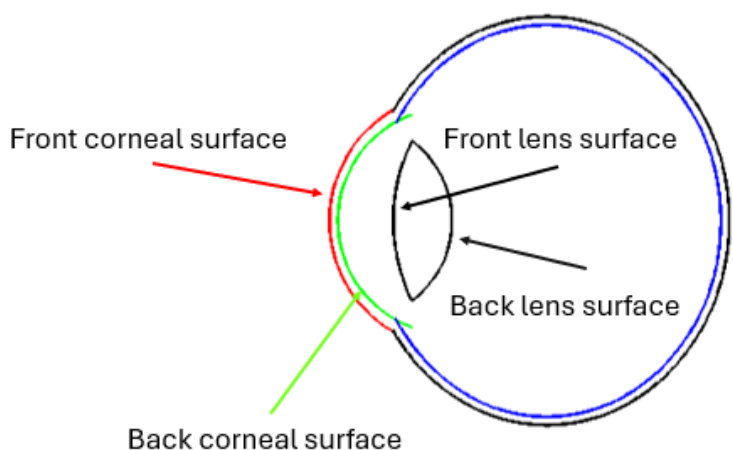

**Figure 1.** The four-surface model of the human eye. The diagram corresponds to spherical surfaces with typical curvature radii.

Basic surface modelling consists of fitting continuous functions to target surfaces, which are expanded as a combination of orthogonal functions [4-6]: $z(x,y) = \sum_{m}^{k} c_m \Phi_{m(x,y)} + \epsilon_k(x,y)$
where coefficients $c_m$ represent the weight of each basic function $\Phi_m(x,y)$, and the completeness of orthogonal sets guarantees that the error $\epsilon_k$ decreases as the number $k$ of terms increases. The coefficients are orthogonal projections of the target onto the "direction" of each basic function and are calculated as definite integrals within the domain of the target surface. The behaviour of the basic functions at the boundary of a disk domain plays a special role in modelling since it might help representing certain features of the target surface. Also, for sampled surfaces, orthogonality is lost, but the mathematics become that of a finite-dimensional linear algebra. While Zernike, Nijboer, Bhatia and Wolf introduced the original Zernike polynomials (see [5] and references therein), the underlying theory left room for exploring several other sets of functions with the key properties of orthogonality and completeness. Among these sets we find polynomials (finite number of terms of powers of the space variables), like Chebyshev, Jacobi, and Zernike polynomials and special functions which can be defined as infinite sums, like Laguerre, Bessel and the Hypergeometric functions [7]. The Zernike polynomials had gained preference amongst researchers and practitioners due to their relation to classical or Seidel aberrations [5], however certain shortcomings appear for applications such as combined surfaces (cornea-limbus-sclera) or membrane deformable mirrors [7,8]. For this special cases, other orthogonal sets have been shown to be better suited [4-7,9]. Figure 2 presents the classical Zernike functions along with the Bessel Circular functions where certain values of the azimuthal order have been selected for specific purposes.

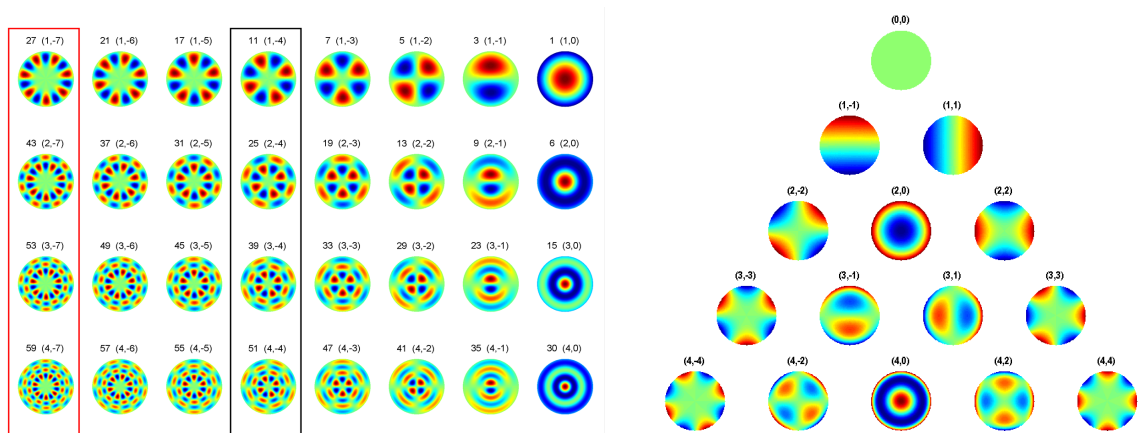

**Figure 2.** A sample of circular orthogonal functions. To the left the Bessel Circular Functions are shown along with the ordering proposed according to Trevino et. al. and organized in terms of azimuthal and radial indices. To the right the classical Zernike Circular Polynomials are arranged according to radial and azimuthal indices.

The anterior corneal surface is mainly modelled for determining keratoconus or for contact lenses fitting among other applications. Zernike decomposition of corneal height data had been proposed by Babcock [10]. However, it was Noll [11] who introduced them for wavefront correction, and Schwiegerling, Greivenkamp, and Miller [12] who launched the Zernike set to be the current standard. Orthogonality of Zernike and related sets is usually defined within a continuous domain, such as the unit disk, corneal topographers based on Placido rings [13,14], produce discrete samples of the corneal data and orthogonality is lost. Efforts were done to overcome this shortcoming by rearranging the sampling distribution, and finally settled by finding the optimal number of samples and amount of Zernike terms needed to represent the cornea with sufficient accuracy [15].

**Current and Future Challenges**

There are several applications for which Zernike polynomials are not optimal. Modelling the crystalline lens, the aberration function and the influence functions of deformable mirrors are a few examples. Although the crystalline lens is a continuous element with variable refraction index (GRIN medium) [16], it has been modelled as a four-surface element and, the latter provides good insight into its optical properties. The main crystalline lens surfaces may be, of course, represented as a finite series of orthogonal functions.

Adaptive optics involves retrieving and correcting the phase aberration of the human eye. The retrieval is done via Hartman-Shack wavefront sensor while the correction is usually done by tuning a deformable mirror. The deformation is attained by shifting actuators to provide the required deformation. Each actuator provides a specific deformation to the mirror's surface. The calibration usually requires a combination of actuators to provide specific Zernike combinations. Although the Zernike representation has worked well, it has been discussed how other types of functions may be better suited for the task.

The peripheral aberrations are measured at angles where the pupil image is seen as an ellipse, and the Zernike polynomials lose the orthogonality property in this domain. There are ways to build orthogonal sets at such conditions taking the original set as a departure point. This process is the Gramm-Schmidt (GS) orthogonalization process [6,17], however there are sets which are naturally orthogonal in the elliptical domain. The natural candidate are the Mathieu functions which are fixed at the boundary, while the Zernikes are free at the boundary which is better suited for this particular problem, so other solutions, such as GS orthogonalization has been applied (see [5] and references therein).

Intra Ocular Lenses (IOLs) characterization [18] is another problem where surface modelling is relevant to enable characterization of their profile and optical properties. Not only determining the best fit given a certain orthogonal set is a problem, but selecting the set that best represents the surfaces is important. Rogel et. al. [4] experimented with several different sets of functions and found that the choice depends on particular features of the target surfaces, which opens the opportunity to test different functions to this complex surfaces. A similar idea applies for Machine Learning, there have been different approaches where the mail goal is to determine the best set of descriptors might be for pre-screening purposes.

**Advances in Science and Technology to Meet Challenges**

Optical modelling has become a cornerstone in addressing the complex challenges faced in modern ophthalmic research and technology. In this context, modelling ocular surfaces, such as the cornea and lens, significantly contributes to the creation of synthetic data sets. Synthetic data is increasingly gaining attention due to its ability to simulate a wide range of biometric variations without the ethical and logistical constraints of real-world data collection. Stochastic eye models are particularly valuable as they generate random biometry data reflecting the natural variability found in the general population, essential for robust optical calculations and the development of personalized medical solutions. Research on synthetic eyes, including those for healthy eyes [18] and keratoconus-affected eyes [19], demonstrates these models' potential to replicate realistic corneal properties and disease conditions. This facilitates the testing and optimization of diagnostic tools and treatments, leading to personalized care. Additionally, advancements in modelling the lens enable the generation of accurate synthetic data for various lens conditions [20,21], further aiding in developing intraocular lenses and other vision correction technologies. Such synthetic data accelerates innovation and ensures a higher standard of safety and efficacy in ophthalmic solutions.

The use of artificial intelligence (AI) techniques to model the corneal surface is becoming increasingly common. However, concerns remain, particularly when studying uncommon or abnormal corneas. Selecting the best set of orthogonal functions can enhance neural network performance, providing a reliable tool for pre-screening potential keratoconus disease. AI plays a pivotal role in enhancing the utility and accuracy of synthetic data in ophthalmology. Machine learning algorithms, in particular, excel at analyzing vast datasets to identify patterns and predict outcomes. When applied to synthetic data, AI refines generated models by learning from both real and synthetic datasets, thereby improving simulation accuracy. This synergy between synthetic data and AI-driven analysis might revolutionize ophthalmic research and technology, leading to more precise diagnostics, personalized treatments, and efficient development of new ophthalmic solutions.

**Concluding Remarks**

Although the idea of surface modelling has been around for several years, the fundamental questions about sampling, orthogonality as related to the high accuracy required for clinical purposes, remain open. The main reason for this is that each surface modelled possesses specific features and therefore demand different sets of orthogonal functions which mirror some of these features, giving the researcher room for improvement.

The automatic assessment of corneas for prescreening purposes has been also around since the explosion of the use of keratometers, however it is only in the last decade that the proper tools have become massively available for researchers and practitioners to test different functions to function as ophthalmic descriptors. The idea is to find a good combination of descriptors that accurately represent the features of the various surfaces and to secure the optimal input for Machine Learning techniques, which also is a vastly unexplored topic.

**Acknowledgements**

Juan Trevino wishes to acknowledge the UPP and Tec de Monterrey for supporting the author's work.

## 13. Peripheral image quality in the human eye

Charlie Börjeson and Linda Lundström,
KTH Royal Institute of Technology, Department of Applied Physics, Stockholm

[ cborje@kth.se; lindafr@kth.se ]

**Status**

Peripheral optics and vision play an important role in daily tasks such as walking and driving, and is crucial for both locomotion and detection. However, the human eye is optimised for central vision, both optically and neurally. The optical image quality gets progressively worse further away from the fovea, with the ganglion cell sampling density decreasing in a similar fashion. The largest aberrations in the fovea (apart from refractive errors) are spherical aberration and longitudinal chromatic aberration [1]. These remain similar in the periphery, but other optical aberrations quickly overtake the effect when going to higher off-axis angles. At first, coma and transverse chromatic aberration will appear, increasing linearly with angle. At even higher angles, oblique astigmatism, that increases approximately quadratically with angle, will start to dominate the image quality. Oblique astigmatism is a result of the apparent asymmetry of a lens when viewing it off-angle, effectively inducing astigmatism that gets worse the steeper the angle. This asymmetry creates two line foci, defining two disparate image planes (see Figure 1). In an emmetropic eye, the retina coincides approximately with the sagittal image plane, where the line foci are oriented radially from the fovea. In the horizontal visual field, the resulting astigmatism therefore corresponds to a correcting negative cylinder lens with axis 90°.

Peripheral refractive errors are rarely measured or corrected, even though already the first reports in the early 1930s by Ferree and colleagues showed that both defocus and astigmatism can be very large [1]. Correcting peripheral refractive errors can be of crucial importance to people with central visual field loss, who rely solely on their peripheral vision. Additionally, they have been found to impact driving and hazard perception also in subjects with normal vision [2,3]. However, in recent years, peripheral optical quality has gained attention in the field of myopia research due to its role in controlling eye growth. Animal studies suggest that not only central defocus, but also peripheral, affect how much the eye grows [4]. Additionally, recently introduced optical myopia control therapies that reduce myopia progression share the common factor that they affect peripheral blur, often aiming to induce relative peripheral myopia [5,6].

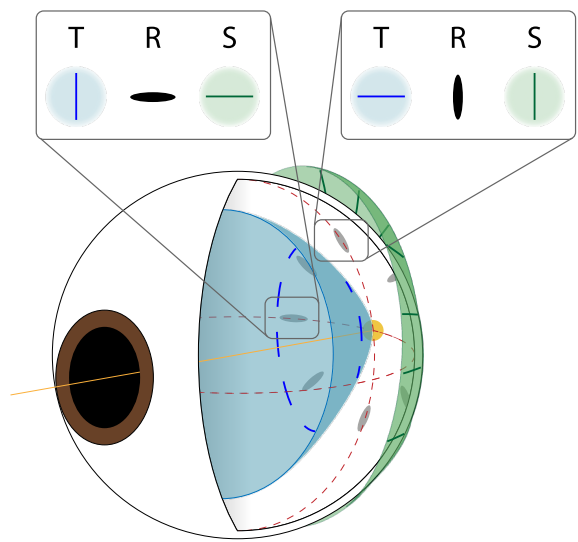

**Figure 1.** Oblique astigmatism over the peripheral field in the human eye. The sagittal (S) image plane is shown in green and has radial line foci. The tangential (T) image plane is shown in blue and has line foci perpendicular to those of the sagittal image plane. The retina (R) often lies in-between the two image planes, closer to the sagittal plane.

**Current and Future Challenges**

Even though there is seldom need to improve the peripheral image quality in the healthy emmetropic eye, optical corrections for ametropic eyes are generally not optimized for peripheral vision. Typically, when the peripheral refractive errors in 20-30° are larger than 1 diopter, they will affect peripheral resolution acuity in both high and low contrast [1]. It is not uncommon that optical correction by means of spectacles, contact lenses, and intraocular lenses induce errors much larger than that [7], [8]. Some manufacturers are currently reviewing the designs of these traditional corrective lenses to address the issues with peripheral vision. This is important for activities such as locomotion and driving where a wide field of view is required. Errors lower than 1 diopter could also be worth correcting, as detection acuity via aliasing is very sensitive to small changes in image quality [1]. This kind of correction could be particularly useful to people with central visual field loss.

The other key question in peripheral image quality research is to understand the function and improve the efficacy of myopia control interventions. Studies have shown that relative peripheral refraction (RPR) may play a role in eye growth regulation and in the development of myopia [4]. Therefore, many of the optical interventions, developed to control myopia, aim to place the peripheral circle of least confusion in front of the retina, i.e. to create myopic RPR and halt further growth. However, RPR is not a single value, but varies over the retina and is dependent on the accommodation state and shape of the eye. For eccentricities of 20° and above, studies have shown that emmetropic eyes usually have negative RPR whereas RPR for already myopic eyes is less negative/more positive [1]. This is thought to be because of a more prolate ocular shape of the myopic eye. With accommodation, simple eye models predict RPR to become more negative because of increased field curvature [9-11]. However, attempts to measure RPR with accommodation are nonconclusive. All of this makes it difficult to quantify and correlate RPR with myopia progression and optical treatment efficacy on group level.

**Advances in Science and Technology to Meet Challenges**

To improve peripheral visual function, it is essential to understand the relation between optical aberrations and vision. One powerful tool used for this is adaptive optics. In an adaptive optics visual simulator, a deformable mirror or a spatial light modulator is used to control what aberrations are present in part of a subject's field of view (centrally or peripherally) [12]. This ensures that all subjects have the same retinal image quality, by eliminating the subjects' individual aberrations. It is then possible to perform vision tests to assess how different optical aberrations affect vision and to what extent they need to be corrected. Adaptive optics can thereby be used to simulate different optical corrections, to understand how they can be improved in terms of their impact on peripheral vision. This can be valuable when developing multifocal designs for combating presbyopia, as these tend to induce additional aberrations. It is especially of use for intraocular lens designs, due to the difficulty of testing new designs in vivo. To make the visual experience as realistic as possible, a large field of view and possibility to allow for head and eye movements is desirable.

To understand the correlation between myopia and RPR, dedicated measurement technologies and long-term studies during ocular growth and during myopia control treatment are needed on individual level. Studies in which accommodation is not paralyzed require additional control of the accommodative state of the eye during the peripheral measurements, for example by using an additional foveal measurement path [13]. For the peripheral measurements, fast scanning devices can be used to quickly evaluate image quality over larger retinal areas [14-19]. However, peripheral refraction through optical myopia control interventions poses additional challenges because of the irregular optical errors of the designs, which make the location of the circle of least confusion ill-defined. These irregular optical errors can also be difficult to pick-up with a conventional wavefront sensor (for an example, see Figure 2) [20]. Hence, other peripheral refraction techniques may be needed, both objective and subjective ones.

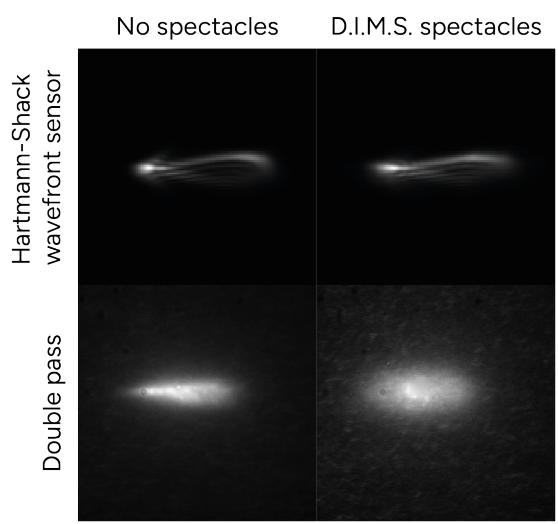

**Figure 2.** Peripheral point spread function (PSF) (at 20°) for a subject wearing either no spectacles or Defocus Incorporated Multiple Segmets (D.I.M.S.) spectacles for myopia control. When measured with a double-pass instrument, there is a clear widening of the PSF for the myopia control spectacles. However, this broadening is not seen with a Hartmann-Shack wavefront sensor.

**Concluding Remarks**

Peripheral image quality is affected by large aberrations such as astigmatism, coma and transverse chromatic aberration. Understanding peripheral optical errors and their correlation with visual function is essential for improved vision for people in all age groups. The two major applications for research on peripheral image quality today are improving optical corrections for peripheral visual tasks and developing efficient optical myopia control interventions.

Peripheral acuity gets worse with increased peripheral optical errors, which are often amplified by conventional spectacles, contact lenses and intra-ocular lenses. This can affect important visual functions necessary for safe locomotion, and is of special concern for people with central visual field loss. Adaptive optics visual simulators can be used to assess new optical designs and to understand how aberrations affect vision.

Myopia progression is affected by peripheral image quality, but the exact mechanisms are not fully understood. Moreover, optical myopia control therapies' effect on peripheral image quality is difficult to measure with conventional instruments. Multi-path and/or scanning devices with control of accommodation are needed to develop more efficient and personalised therapies.

**Acknowledgements**

The authors currently have funding for research within myopia by the Swedish Research Council (No. 2019-05354 and No. 2023-05428) and within vision for elderly by the European Union through the MSCA Doctoral Network ACTIVA (101119695).

## 14. Optical aberrations of the human eye

Seung Pil Bang, MD, PhD[1], Geunyoung Yoon, PhD[2]

[1]Department of Ophthalmology, Keimyung University Dongsan Medical Center, Daegu, South Korea
[2]College of Optometry, University of Houston, Houston, TX, USA

[gibong87@dsmc.or.kr ; gyoon2@central.uh.edu]

**Status**

Over the past several decades, the study of monochromatic aberrations in the human eye has been a focal point in physiological optics. Early methods such as the Scheiner double pinhole [1], the Foucault knife edge [2], and the crossed-cylinder aberroscope [3] were attempted, but found to be labor-intensive and difficult to analyze. In the 1980s, the rise of corneal refractive surgery led to increased patient complaints about glare and visual discomfort, sparking significant interest in measuring ocular aberrations within ophthalmology and vision science. This demand for precise quantification of the ocular aberrations prompted the development of the highly effective Shack-Hartmann sensor [4,5], which represented a notable technological breakthrough. Subsequent advancements introduced various techniques, including the ray-tracing aberrometer [6], the curvature sensor [7], and the pyramidal sensor [8]. These sensors primarily serve three purposes: (1) rapid and objective quantification of ocular aberrations, (2) understanding the impact of the aberrations on visual performance, and (3) facilitating vision correction through adaptive optics [9], lasers [10] and ophthalmic lenses [11]. Technological advances have expanded their clinical applications to other areas including tear film and dry eye [12], characterization of ocular conditions like keratoconus [13] and cataracts [14], presbyopia correction [15], and myopia control [16].

**Current and Future Challenges**

Current ocular wavefront sensing technology presents several technical challenges. One primary limitation is the spatial resolution imposed by the discrete sampling of the wavefront. For example, the size of lenslets in a Shack-Hartmann sensor can lead to spatial under-sampling [17], causing variations within each lenslet to be averaged and lost. Additionally, intraocular scatter induced by cataracts or scars can degrade spot quality, increasing errors in estimating the wavefront slope. Ocular aberrations also vary over time due to factors such as fluctuations in the tear film's thickness and refractive index, accommodation, and fixational eye movements [12]. The speed at which these changes can be traced is limited by the sensitivity of the CCD and CMOS sensors at the wavelength of light used. The very low light levels (0.01-0.001% of input power) from the eye further complicate this challenge. Trade-offs exist between spatial and temporal resolution in ophthalmic applications. Additionally, it is challenging to quantify peripheral aberrations, which provide important clues about how peripheral optics influence emmetropization and refractive error development [18,19]. A few scanning wavefront sensors have been proposed, but improvements in compactness and portability are needed. Another challenge lies in wavefront reconstruction methods. Another challenge lies in wavefront reconstruction methods. While modal wavefront reconstruction (using Zernike polynomials) is common in commercial sensors and decomposes different types of aberrations, it acts as a low-pass spatial filter, removing potentially important high spatial frequency wavefront information. The zonal reconstruction algorithm [20] can overcome this challenge and needs to be refined and implemented in addition to the modal algorithm.

Although a complete assessment of the optical imperfections of the eye provides insights into the image quality formed on the retina, the complex post-receptoral and neural processing contributing to subjective perception is of significant scientific interest. One intriguing finding is that the neural system can reduce perceived optical blur through long-term adaptation to the individual eye's habitual optical profile [21,22]. Further study is needed to identify the mechanisms by which visual experience with specific ocular optics alters our perception. This research will also improve our ability to predict subjective refractive errors from measured aberrations [23]. Selecting the reference axis along which the eye's aberrations are measured can significantly impact the outcomes of advanced vision corrections with lasers and ophthalmic lenses [24]. Clinical accessibility to wavefront sensors is a significant challenge due to the lack of low-cost, stand-alone aberrometers available commercially. This limitation hinders clinicians' ability to diagnose patients' visual problems and enhance corrective outcomes.

**Advances in Science and Technology to Meet Challenges**

Advances in sensor technologies [25] have enabled increased spatial resolution of Shack-Hartmann wavefront sensors by magnifying the eye's pupil onto the lenslet array plane with a 4-f telescope system [17]. This effectively reduces the sampling steps on the eye's pupil by the magnification factor. Other phase imaging techniques, such as quadriwave lateral shearing interferometry [26] and phase diversity/retrieval [27], also offer the potential for achieving higher sampling resolution. These technologies will eventually allow for the direct quantification of intraocular scatter. However, due to the relatively low quantum efficiency (QE) of silicon-based sensors at near-IR wavelengths, the methods mentioned above require higher power of the input laser closer to the maximum permissible exposure (MPE), raising eye safety concerns. To overcome this issue, using an indium gallium arsenide (InGaAs) sensor, which has excellent QE (> 80%) in the short-wave infrared band (>900 nm), is a viable solution. This newly available sensor offers benefits including higher MPE at longer wavelengths and a substantially faster frame rate, making it possible to measure dynamic ocular aberrations at significantly higher speeds [28]. Moreover, the ability to use a longer wavelength of light that is completely invisible to the eye opens new possibilities for testing visual functions across a wide range of the visual field without visual disturbances from the light source.

Optical tools such as adaptive optics and free-form ophthalmic lenses based on ocular wavefront sensing are key to understanding the interaction between the eye's optics and visual perception. These tools allow researchers to bypass optical imperfections, enabling the evaluation of neural functions and the induction of specific optical profiles for comparison of neural processing across different individuals. Along with advancements in these techniques, developing innovative visual psychophysical paradigms and optical imaging modalities is critical for revealing mechanisms related to various aspects of human vision. Ongoing progress includes creating more sophisticated visual quality metrics that represent real-world visual tasks, incorporating factors such as spatial frequency, contrast, color distribution in natural scenes, retinal sampling frequency, motion, and binocularity.

**Concluding Remarks**

Ocular wavefront sensing technology has revolutionized our understanding of the eye's optical imperfections and how to correct them to improve visual performance. Scientific and clinical research findings are now more objective and precise, offering valuable insights into how human vision works in relation to the eye's optical characteristics. Addressing these challenges will enhance our ability to characterize the eye's optics both spatially and temporally, under various ocular conditions. Successfully translating these advancements into practical vision correction will enable individuals to achieve optimal visual quality customized to their needs, significantly improving their quality of life.

**Acknowledgements**

Supported by United States National Eye Institute (grant number: R01 EY014999 and R01 EY034151).

Advances in Visual and Physiological Optics Roadmap, *Journal of Optics*[16] Ji, Q., Yoo, Y. S., Alam, H., and Yoon, G. (2018). Through-focus optical characteristics of monofocal and bifocal soft contact lenses across the peripheral visual field. *Ophthalmic Physiol. Opt.*, **38**(3), 326-336.

[17] Bang, S. P., Jung, H., Li, K. Y., and Yoon, G. (2024). Comparison of modal and zonal wavefront measurements of refractive extended depth of focus intraocular lenses. *Biomed. Opt. Express*, **15**(3), 1618-1629.

[18] Pusti, D., Kendrick, C. D., Wu, Y., Ji, Q., Jung, H. W., and Yoon, G. (2023). Widefield wavefront sensor for multidirectional peripheral retinal scanning. *Biomed. Opt. Express*, **14**(8), 4190-4204.

[19] Jaeken, B., Lundström, L., and Artal, P. (2011). Fast scanning peripheral wave-front sensor for the human eye. *Opt. Express*, **19**(8), 7903-13.

[20] Panagopoulou, S. I., and Neal, D. R. (2005). Zonal matrix iterative method for wavefront reconstruction from gradient measurements. *J. Refract. Surg.*, **21**(5), S563-9.

[21] Artal, P., Chen, L., Fernandez, E. J., Singer, B., Manzanera, S., and Williams, D. R. (2004). Neural compensation for the eye's optical aberrations. *J. Vis.*, **4**(4), 281-7.

[22] Sabesan, R., and Yoon, G. (2009). Visual performance after correcting higher order aberrations in keratoconic eyes. *J. Vis.*, **9**(5), 6 1-10.

[23] Thibos, L. N., Hong, X., Bradley, A., and Applegate, R. A. (2004). Accuracy and precision of objective refraction from wavefront aberrations. *J. Vis.*, **4**(4), 329-51.

[24] Bang, S. P., Lyu, J., Ng, C. J., and Yoon, G. (2022). Visual axis and Stiles–Crawford effect peak show a positional correlation in normal eyes: A cohort study. *Invest. Ophth. Vis. Sci.*, **63**(11), 26-26.

[25] Eddaif, L., and Shaban, A. (2023). Chapter 2 - Fundamentals of sensor technology. In A. Barhoum and Z. Altintas (Eds.), *Advanced Sensor Technology* (pp. 17-49). Elsevier.

[26] Baffou, G. (2021). Quantitative phase microscopy using quadriwave lateral shearing interferometry (QLSI): Principle, terminology, algorithm and grating shadow description. *J. Phys. D-Appl. Phys.*, **54**(29), 294002.

[27] Bonaque-Gonzalez, S., et al. (2021). The optics of the human eye at 8.6 microm resolution. *Sci Rep*, **11**(1), 23334.

[28] Jung, H. W., Bang, S., Pusti, D., and Yoon, G. (2023). High-speed Shack-Hartmann ocular wavefront sensor using an InGaAs sensor. *Invest. Ophth. Vis. Sci.*, **64**(8), 2888-2888.62

## 15. Intraocular pressure of the human eye

Karol Karnowski[1,2], Bartlomiej J. Kaluzny[3], Ireneusz Grulkowski[4]

[1] International Centre for Translational Eye Research, ul. Skierniewicka 10A, 01-230 Warsaw, Poland
[2] Institute of Physical Chemistry, Polish Academy of Sciences, ul. M. Kasprzaka 44/52, 01-224 Warszawa, Poland
[3] Department of Ophthalmology, Collegium Medicum, Nicolaus Copernicus University, ul. M. Curie Skłodowskiej 9, 85-094 Bydgoszcz, Poland
[4] Institute of Physics, Faculty of Physics, Astronomy and Informatics, Nicolaus Copernicus University, ul. Grudziadzka 5, 87-100 Toruń, Poland

[kkarnowski@ichf.edu.pl ; b.kaluzny@cm.umk.pl ; igrulkowski@fizyka.umk.pl]

**Status**

The intraocular pressure (IOP) stands as a fundamental parameter that maintains the eyeglobe's shape and enables the physiological functions of ocular tissues. Typically ranging from 10 to 21 mmHg, normal IOP is essential for preserving the eye's optical properties. The precise regulation of IOP is determined by the interplay between the production and outflow of aqueous humor within the posterior and anterior chambers. Elevated IOP is a primary risk factor for glaucoma, a group of optic neuropathies that are a leading cause of irreversible blindness worldwide. Beyond glaucoma, IOP may reflect pathophysiological changes that indicate the onset of conditions such as uveitis, pigment dispersion syndrome, or angle-closure crisis.

The measurement and interpretation of IOP extend far beyond mere numbers. In glaucoma management, IOP serves as both a diagnostic criterion and a therapeutic target. While not all patients with ocular hypertension develop glaucoma, reducing IOP remains the only proven strategy to slow the progression of glaucomatous damage. This makes accurate IOP assessment indispensable for initiating timely interventions that can preserve vision.

Throughout history, the need to measure IOP accurately has driven innovations, leading to a variety of techniques [1,2]. Goldmann Applanation Tonometry (GAT) has long been regarded as the gold standard. It estimates IOP by gently flattening a small area of the cornea using a prism attached to a slit lamp microscope. While GAT offers reliable results, the need for topical anesthesia introduces potential risks, such as corneal abrasions or infections. Additionally, variations in central corneal thickness and biomechanical properties of the cornea can influence readings, necessitating careful interpretation.

Non-contact methods, like the air-puff tonometer, offer a touch-free alternative by using an air-pulse to deform the cornea. This approach reduces the risk of infection and enhances patient comfort; however, it may sacrifice some accuracy compared to GAT and is sometimes not well tolerated by patients. The Ocular Response Analyzer (ORA) directly measures light reflected from the cornea during air-puff-induced applanation. The Corvis ST (Corneal Visualization Scheimpflug Technology) combines air-puff tonometry with high-speed Scheimpflug imaging to visualize corneal response. Both approaches go a step further by also assessing cornea's biomechanical properties.

Dynamic Contour Tonometry (DCT) seeks to overcome corneal influence by matching the device's contour to that of the cornea, aiming for IOP readings less affected by corneal properties. Represented by the Pascal tonometer, DCT is a slit-lamp-mounted unit with a pressure sensor tip on a contact area and enables the measurement of IOP and ocular pulse amplitude.

Handheld devices bring flexibility to the clinical setting, allowing measurements in various patient positions and circumstances. These devices are particularly useful for pediatric patients or those who have difficulty with traditional slit lamp examinations. Tono-Pen is a handheld device combining elements of the applanation and indentation methods. Rebound tonometry, such as a series of iCare tonometers, measures the deceleration of a small probe that bounces off the cornea and has become a popular alternative to GAT.

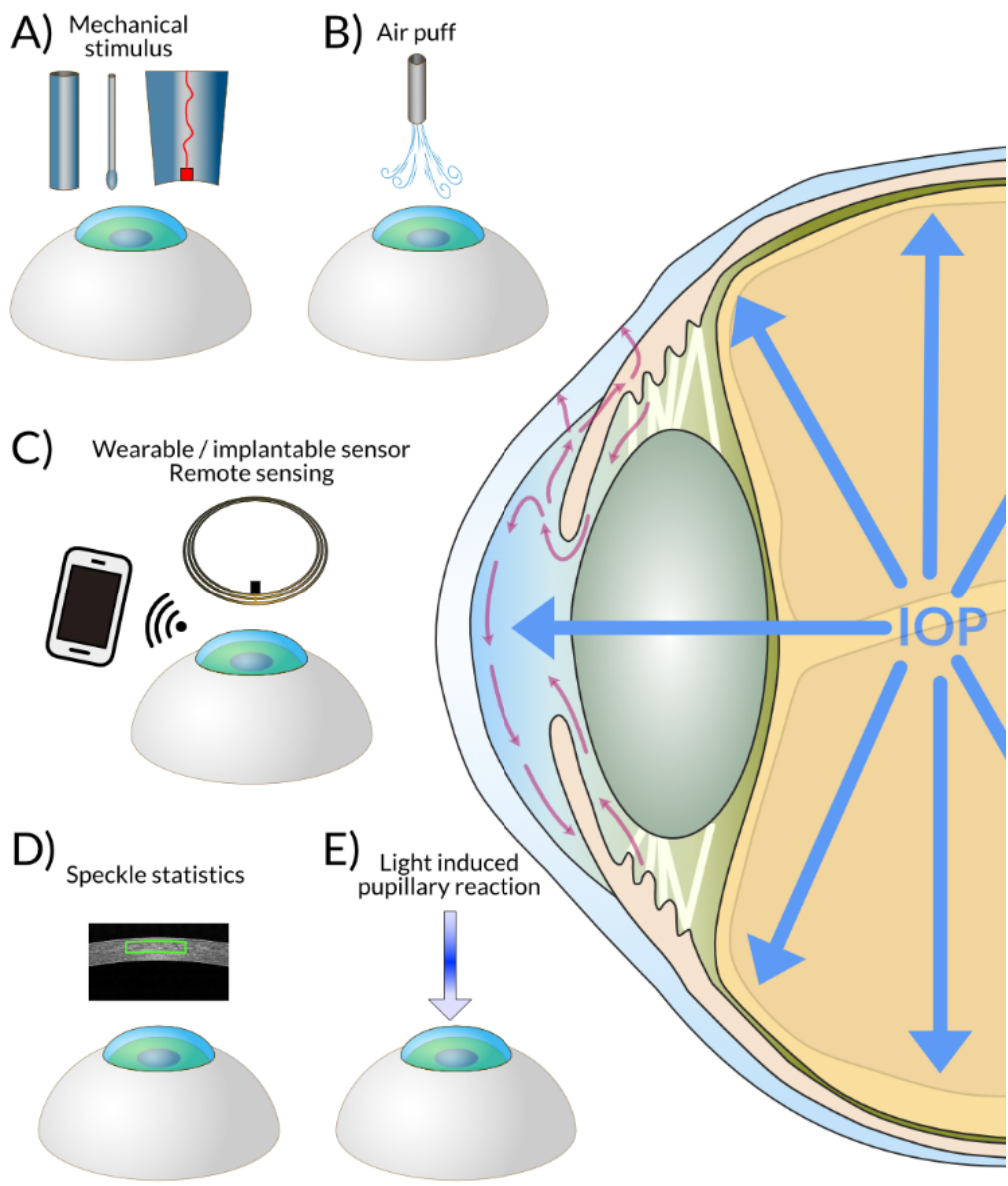

**Figure 1.** Recent and Emerging IOP Measurement and Monitoring Methods. On the right, a cross-sectional view of the human eye displays the aqueous humor circulation pathways indicated by red arrows. Intraocular pressure (IOP) affects all structures of the eye. A) Mechanical stimulus methods (e.g., Goldmann tonometer, iCare, or DCT). B) Non-contact methods based on air puff stimuli. C) Wearable or implantable pressure sensors enabling remote IOP monitoring. D) IOP estimation based on statistical analyses of corneal cross-section images. E) IOP values extracted pupillary reaction to light stimulus.

**Current and Future Challenges**

Clinically available techniques for estimating IOP are based on indirect measurements, which generates difficulties in the proper interpretation of the results. Fundamental challenges in IOP estimation stem from the inherent biomechanical complexity of the eye, which introduces variability in the IOP results. Differences in **corneal geometry (**thickness, curvature), hydration, corneal biomechanics (rigidity), and cardiac cycle have been identified as factors leading to inaccuracies in IOP readouts [3]. Thicker corneas often result in an overestimation of IOP, while thinner corneas may yield lower results. Diurnal **fluctuations in IOP** also need to be taken into account during the assessment of a patient's glaucoma risk. Additionally, the IOP measurement procedures is impacted by the limitations of existing tonometry technologies, which often require **direct contact** with the eye or the use of anesthetics. This generates discomfort, increases infection risks, and contributes to patient anxiety, particularly among children and the elderly. Several studies demonstrated another complexity, namely the **lack of consistency between different tonometry devices, which** often provide different IOP measurements [4]. Consequently, it is challenging to standardize the results across clinics or in longitudinal studies.

Future challenges in IOP measurement are driven by the characteristics of the ideal tonometer. The clinical practice indicates that an ideal tonometer should combine features such as reliability, patient comfort, ease of use, and portability. It should provide reliable readings across different patients accounting for variations in corneal thickness, curvature, and biomechanics. The device should minimize or eliminate physical contact with the eye or use stimuli better tolerated by patients. Furthermore, the ideal tonometer should have high inter- and intraoperator reproducibility, allowing less specialized personnel to perform the measurements objectively. The ideal technology should also accommodate the need for portability, automated measurements, and compatibility with data-sharing resources. Additional future challenges include the development of robust, cost-effective solutions to increase patients' access to new technologies in developing regions.

**Advances in Science and Technology to Meet Challenges**

The challenges listed above underscore the need for continued innovation and refinement in IOP evaluation methods. In response, the market has been upgrading existing devices and developing novel solutions for IOP assessment.

Non-contact tonometry now provides more advanced quantitative analysis of air-puff-induced corneal deformation, allowing for a better understanding of the relation between IOP and corneal biomechanics. Implementation of new versions of software takes into account the compensation of factors skewing IOP measurements, which enhances the diagnostic capability of air-puff tonometers (e.g., Goldmann-correlated IOP and corneal-compensated IOP in ORA, and biomechanically-corrected IOP in Corvis ST).

Old tonometry concepts are also being realized in low-cost systems on widely available smartphones [5, 6]. Air-puff stimulation can be combined with other modalities, such as optical coherence tomography (OCT), which enables high-resolution imaging and more precise identification of corneal surfaces [7]. Recent advances in OCT technology allow for improving imaging capabilities (e.g., simultaneous horizontal and vertical OCT B-scan pair, optical biometry) for a more comprehensive analysis of ocular tissue response to stimuli [8,9].

The development of rebound tonometers addresses the challenges of IOP measurement, enabling wider use of these devices. Improved versions of iCare, iCare PRO, and iCare IC200 allow for fast and more precise IOP measurement when the patient is in a supine or upright position. Recently introduced new generation, iCare HOME, is the first hand-held device offering self-tonometry at home during regular daily activities [10,11].

Perhaps most intriguing are innovations like implantable pressure sensors and smart contact lenses. Those emerging solutions utilize different types of sensors capable of capturing IOP fluctuations throughout the day that single measurements might miss. Smart contact lenses or intraocular lens implants embedded with micro-sensors provide a non-invasive means to monitor IOP changes with a cost-effective smartphone-based detection system [12-14]. The development of wearable and implantable technology and self-tonometry promises to transform patient care [15,16].

Innovative drug delivery systems are also leveraging IOP as a trigger for therapeutic action. Devices that release medication in response to elevated IOP could provide targeted treatment for glaucoma patients, ensuring timely intervention when it is most needed [15,17].

Recent alternative IOP measurement methods aim to overcome existing challenges by eliminating mechanical stimulation. One promising prototype, tested on 31 eyes without elevated IOP, achieved results comparable to GAT by measuring pupil response using a near-infrared camera and pulsed illumination at the wavelength of 482 nm, though it requires patient-specific calibration [18]. Additionally, OCT speckle analysis has shown potential in ex vivo porcine eyes with good correlation with controlled IOP, but in vivo human tests using air-puff tonometry references exhibited poor correlation, indicating the need for further research [19].
Furthermore, understanding how IOP influences ocular biomechanics contributes to broader insights into eye health. Investigations into myopia progression, for instance, consider the role of IOP in scleral remodelling [20]. By examining these correlations, scientists hope to uncover new strategies for managing refractive errors and preventing vision loss.

Integrating artificial intelligence and machine learning into IOP data analysis offers another avenue for innovation. By recognizing patterns and predicting disease progression, these tools could enhance early detection of glaucoma and optimize therapeutic interventions [21].

**Concluding Remarks**

Intraocular pressure remains a cornerstone of ophthalmic practice, its measurement and management are essential for preserving vision and preventing blindness. While traditional techniques have provided a foundation for care, the challenges they present highlight the need for continued innovation. Emerging technologies offer exciting possibilities for more accurate, continuous, and patient-friendly IOP monitoring, paving the way for personalized medicine and improved outcomes.
Global health initiatives recognize the importance of accessible eye care. Developing affordable, user-friendly tonometers for use in under-resourced areas could make a significant impact on the global burden of glaucoma-related blindness. Telemedicine platforms, enhanced by smartphone-based technologies, may extend the reach of ophthalmic services, bringing essential care to remote communities.

**Acknowledgements**

The authors gratefully acknowledge support from the National Science Center (IG; 2021/43/I/NZ5/03328), International Research Agendas programme of the Foundation for Polish Science co-financed by the European Union under the European Regional Development Fund (KK; MAB/2019/12 and FENG.02.01-IP.05-T005/23).

# 16. Varying optical power design of progressive spectacles, contact, and intraocular lenses

Sergio Barbero

Instituto de Óptica (CSIC), Serrano 121, Madrid, Spain

[ sergio.barbero@csic.es ]

**Status**

The degradation of the accommodation capabilities of the crystalline lens with age, i.e. presbyopia, causes blurred images when viewing objects at different distance locations. When presbyopia combines with far-vision ametropia, an optical element providing varying optical power, either through continuous extended-depth-of-focus (EDOF) or discontinuous multifocal (MF) variation, is required. Bifocals spectacles, already known in Benjamin Franklin's times (late 18th century), one of his inventors, were the first optical elements providing multifocality. However, it was not only until the twentieth century when with the advent of, on the one hand, contact, and intraocular lenses, and, on the other, progressive addition lenses, that optical technology offered multifocal solutions covering not only far and near vision but also intermediate distances. Moreover, in recent times, multifocal solutions are also being used for other visual functionalities, such as myopia control [1]. In ophthalmic lenses, these designs resort to geometric, diffractive, or a combination of both effects. The goal consists of distributing light intensity along different foci (MF) or an extended region (EDOF). A surface with non-constant mean curvature is required when a refractive element is employed. Here, we focus on the optical design challenges set by EDOF and MF solutions; hence, our analysis excludes procedures based on reshaping human eye tissues, such as corneal reshaping through orthokeratology, or the ambitious goal of restoring accommodation. Another way to classify varying optical power solutions for vision is by considering the underlying interaction between the visual system and the optical element. Basically, there are two options. In one case, when the eye looks at a fixed distance, the optical element, either an intraocular (IOL) or contact lens (CL), provides, on the retina, an in-focus image with superimposed blurred images formed by other object planes. Afterwards, the neural system is supposed to be able to select the focused image for each visual target, filtering the blurred signals from the other foci that are perceived as ghost images or halos. This is called the simultaneous vision principle, illustrated in Figure 1.

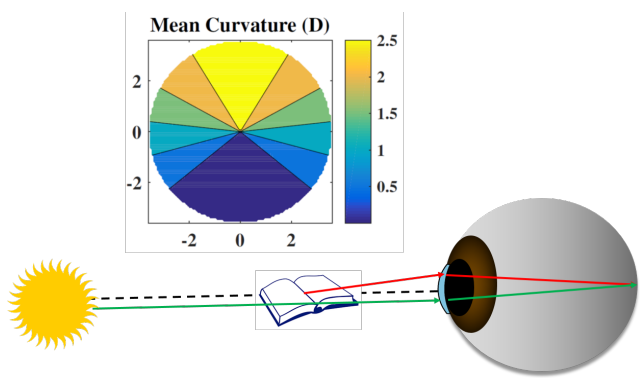

**Figure 1.** Scheme showing how the simultaneous vision principle works illustrated with a contact lens. Rays from a far object (green arrow emerging from the sun) passing through one region of the contact lens converge onto the retina. Conversely, rays from a near object (red arrow emerging from the book) only converge onto the retina if they pass through other parts of the contact lens. Multifocality is achieved by employing a surface with variable mean curvature, as illustrated in the upper figure.

The second case, employed in progressive addition spectacles (PALs, for short), is sometimes called the alternating principle. A surface, or more rarely the combination of two, is designed to contain a spatially varying mean curvature. The eye scans, sequentially on time, each viewing area of the PAL searching for the required optical power. There are two clear zones: far-distance vision (center of the lens) and near-view (low nasal portion) and in between these two zones, the power varies progressively. See Figure 2.

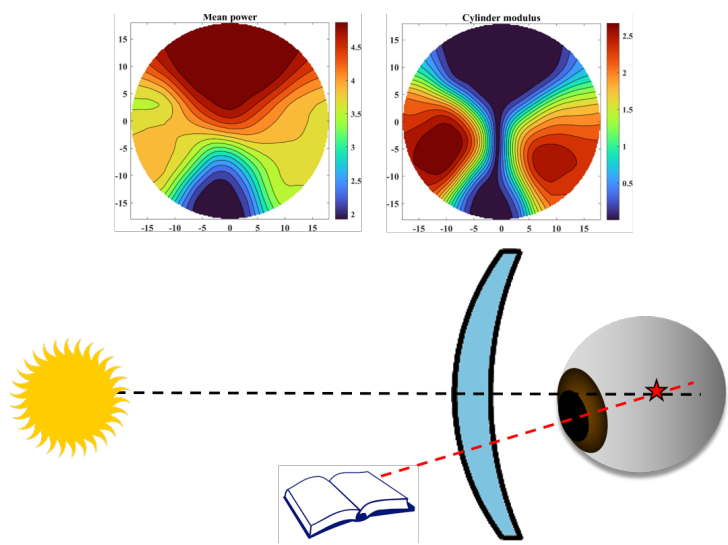

**Figure 2.** Scheme showing how the alternating vision principle works. The eye changes its gaze direction scanning the different parts of the PAL surface searching for the required optical power. The upper figure shows mean curvature and cylinder distribution maps of a realistic PAL.

Nevertheless, the classification above admits some exceptions; for instance, some rigid contact lenses are designed to permit slight centering variations with the viewing distance, thus working under the alternating principle [1].

**Current and Future Challenges**

The greatest challenge in PAL design is to find a smooth progressive mean curvature surface (providing the prescribed optical power) but with a controlled amount of induced astigmatism. Differential geometry theorems make the presence of astigmatism unavoidable [2] because, within a smooth Euclidean surface, zero astigmatism is only possible at isolated points or lines. Minkwitz's theorem establishes that along an umbilical (zero astigmatism) line of a smooth surface with a rate of change of mean curvature given by $k_r$, the cylinder along the orthogonal direction to that line increases twice as quickly as $k_r$. Recently, it has been derived a generalization of Minkwitz's theorem that provides the exact magnitude of the cylinder at any arbitrary point as a function of the geodesic curvature and the ratio of change of a principal curvature along a principal line [2]. Although designing PALs admits several approaches, the key to the majority is to define a cost function that includes different weighted targets, which is subsequently minimized [3]. Typically, the cost function prescribes mean power and cylinder. Moreover, current designs consider that both mean power and cylinder depend, not only on the progressive surface differential properties but also on the angle of incident of the ray onto the surface, i.e. the gaze direction [3]. Besides mean power and cylinder, other targets to the cost function may be considered, for instance, power and cylinder variation, distortion, or even higher-order aberrations [3]. Concerning IOLs and CLs, there are several types of designs: multifocal, designed to provide two or more focuses with separated optical zones, bifocal (multifocality restricted to two focuses), and varifocal power (EDOF), where the refractive power varies progressively with more or fewer smoothness levels. For a long time, these lenses have been restricted to be rotationally symmetric; either concentric designs (segmented discrete rings of alternating power) or aspheric designs (described by a conic plus some extra aspherical coefficients). However, in recent times, the improvement in the manufacturing capabilities of free-form lenses and phase masks, and the progress in their theoretical design, have made possible non-rotationally symmetric designs; starting with the work of Kołodziejczyk et al. [4]. These exhibit two major advantages concerning the former. First, major robustness to pupil changes, since the varying size of the pupil due to light intensity variation, accommodation, or aging limits the amount of light reaching the retina, thus affecting the visual contrast and the multifocality performance. Second, the capacity to provide a higher light concentration in the desired regions [5]. Besides pupil size, correct lens centering and the interaction between the optical element and subjects' aberrations are critical factors; particularly, in soft contact lenses, where decentring is normally induced in the inferior and temporal direction due to gravity and eyelids. Some designs have explicitly considered the eye-lens aberrations coupling, either using a generic pseudophakic eye model [6] or some personalized anatomic data [7].

An essential issue in simultaneous vision solutions is to select the mathematical metric, required in the design process, that better matches the visual experience. Typically, the depth-of-focus or multifocality of the eye is estimated from objective optical computations using various retinal image quality metrics [8]. These combine the purely optical effects of the optical system with an estimate of the subsequent neural process. The boom in experimental visual simulators, with the emergence of techniques allowing spatial control over the phase of light beams, allows direct analysis of which designs offer a better visual experience. A subsequent design procedure requires obtaining the geometry of the lens from the optimal wavefront at the pupil plane [9].

**Advances in Science and Technology to Meet Challenges**

Ophthalmic lenses are almost universally made of bulk lenses that change light through refraction and/or diffraction. However, the spring up of flat optical-element technologies, using metamaterials and metasurfaces, offers exciting potentials in multifocality applications for vision thanks to, on the one hand, their thin and lightweight structures and, on the other hand, their ability of high-precision phase manipulation employing nanostructures at a subwavelength scale. Two design principles applied are the magnetic dipoles and the Pancharatnam-Berry phase, or geometric phase (GP), which generates a varying phase through a closed path [10]. The number of foci and the amount of energy light distribution can be controlled by setting the structure, number, and thickness of stacked GP layers. Some MIOL designs already employ a geometric GP structure. Particularly, Na et al. [11] and Lee. et al. [12] proposed combining the GP structure with a thick refractive lens. Nevertheless, the scaling up to mass production of ophthalmic lenses including metamaterials still poses some significant hurdles because of the requirement of high-precision fabrication techniques, such as electron beams or nanoimprint lithography [13]. Another potential disruptive technology, which, although ultimately oriented to accommodation restoration, could also be used in multifocality or even as a hybrid technology, is the integration of electrically tunable liquid crystal lenses (TLCL). These devices explore the capacity to modify the refractive index of liquid crystals by applying variable electric fields. However, the great challenge in TLCLs for ophthalmic solutions is to detect the object locations for different viewing distances and send that information to the TLCL. Some examples of the application of this technology are: IOLs [14], CLs [15], and spectacles [16]. In PAL design, one of our research interests lies in finding the most convenient cost function, which in recent times has been accepted to be, to some degree, patient-dependent. Therefore, a lot of effort is being put into feeding the personalized designs with individual user parameters such as the patient's relationship between head and eye movements (gaze dynamics), the daily activities of the wearer, the influence of distortion on perception [17], or the binocularity function. Theoretically, once the cost function is properly configured, it would be highly convenient to prove the existence and uniqueness of its solution. This goal has been pursued by [18] considering the classical functional that only combines mean power and cylinder as targets. Here, it is worth noticing the connection between this function and Willmore's; a very active field of research in mathematics [19]. Moreover, this connection may also be used for designing astigmatism-free transition zones in segmented multifocal designs [19]. Finally, PAL designs would benefit from any advances in the progress of numerical algorithms that improve the ability to reach optimal solutions [20].

**Concluding Remarks**

Due to the overall aging of the world's general population, the social impact of presbyopia correction is becoming increasingly important. Both theoretical and technological challenges make the technologies of multifocality and extended-depth-of-focus a very active field in visual optics. Moreover, some disruptive new technologies that have emerged in recent times, such as free-form high-performance manufacturing or meta lenses, to name a few, exhibit exciting potential for the near future.

**Acknowledgements**

This work was supported by grant PID2020-113596GB-I00 from the Spanish Ministerio de Ciencia e Innovacion.

## 17. Intraocular lenses and peripheral vision

Pablo Artal and Juan Tabernero
Laboratorio de Optica, Universidad de Murcia, 30100 Murcia, Spain

[ juant@um.es ; pablo@um.es]

**Status**

In a natural phakic eye, the crystalline lens is a thick lens with a gradient refractive index that supports high-quality imaging over a broad visual field [1]. Traditional intraocular lenses (IOLs), however, are primarily designed to optimize central or foveal vision, often neglecting the quality of vision in the peripheral retina. Most IOLs feature a thin, biconvex design to facilitate easier implantation, but this simplicity comes at a cost: a degraded image quality in the peripheral retina compared to that provided by the natural crystalline lens.

While the limitations of IOLs in peripheral vision have been known for some time [2], they have received little clinical attention, based on the assumption that peripheral optical degradation would remain within the retinal and neural tolerance limits. This would imply a limited impact in peripheral and functional vision. However, recent research challenges this view. Jaeken et al. [3] demonstrated that patients implanted with biconvex IOLs experienced significantly more astigmatism and defocus in their peripheral vision compared to the natural lens in their fellow eye. Furthermore, Togka et al. [4] found that this optical deterioration reduces contrast sensitivity in the peripheral visual field, affecting the patient's ability to detect objects and changes in contrast outside the central vision area. Reduced peripheral visual quality after cataract surgery can have considerable implications for a patient's quality of life [5]. Peripheral vision is essential for many daily tasks, including navigation, recognizing objects in the surrounding environment, tracking multiple objects simultaneously, and planning eye movements. Degraded peripheral optics can impact fundamental activities like navigating stairs [6] and driving [7], where awareness of one's surroundings is crucial. Moreover, compromised peripheral vision is associated with an increased risk of falls, underscoring its importance in maintaining spatial orientation and overall safety.

These findings highlight the need for innovation in IOL design that accounts for peripheral as well as central vision. Some recent developments have focused on addressing these issues by modifying the design of IOLs to improve peripheral image quality, such as incorporating non-biconvex shapes or adjusting the refractive index profile. By better mimicking the natural crystalline lens, these advanced IOLs could help restore a more natural, seamless visual experience after cataract surgery. Improving peripheral optics in IOLs not only enhances central and peripheral vision balance but also holds potential for significantly improving patients' functional abilities and quality of life.

**Current and Future Challenges**

These findings of suboptimal peripheral optics and their visual consequences in pseudophakic patients prompted the development of an alternative intraocular lens (IOL) designed to preserve high-quality peripheral optics through an inverted meniscus design [8]. These lenses are intended to emulate the peripheral optical performance of the natural crystalline lens. In this design, the posterior surface curvature radius is smaller than the anterior surface curvature, with this ratio varying according to the IOL's power. Figure 1 illustrates an example of this lens type (ArtIOL, Voptica SL, Murcia, Spain).

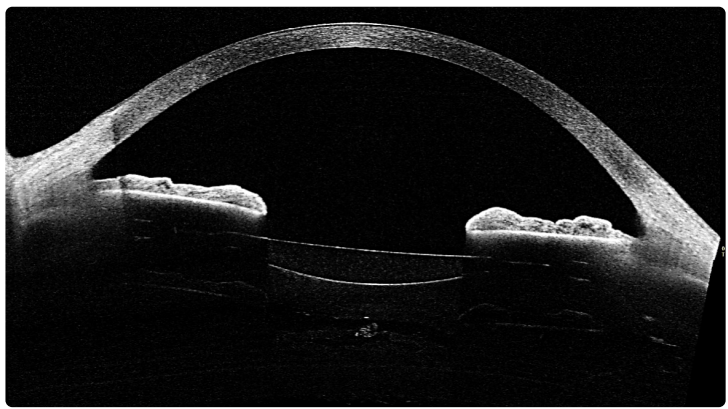

**Figure 1.** Anterior OCT image of a patient implanted with an inverted meniscus IOL.

**Advances in Science and Technology to Meet Challenges**

The inverted meniscus lenses have already been implanted in a significant number of patients, demonstrating notable improvements in both image quality and peripheral contrast sensitivity. Figure 2 illustrates the average peripheral refraction across a cohort of patients with standard biconvex lenses compared to those with inverted meniscus lenses [9]. In these patients the impact of a better optics was a better contrast sensitivity in the periphery.

On the other hand, standard biconvex IOLs tend to introduce eccentricity-dependent shifts in the visual field, a distortion that appears to be minimized with inverted meniscus IOLs. We compared pre- and post-operative fundus images from cataract patients to assess shifts in retinal landmark positions induced by IOL implantation at different eccentricities. Results indicated that these displacements were considerably less pronounced with inverted meniscus IOLs than with traditional biconvex lenses. This finding suggests that standard IOLs may disrupt the world-to-retina mapping, potentially leading to distortions in the perceived positions of peripheral objects. Such visual inaccuracies could increase the likelihood of spatial errors, potentially resulting in misjudgments or accidents, particularly in complex visual environments. However, it is possible that neural adaptation processes could gradually restore the normal perception of the visual field, correcting for these distortions over time. Further studies are needed to fully understand the extent of this adaptation and its implications for patient safety and quality of life.

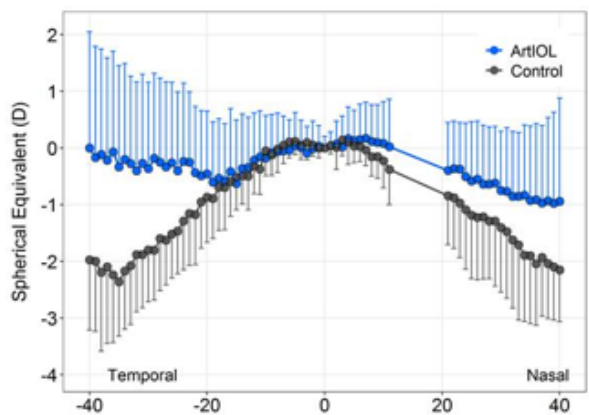

**Figure 2.** Peripheral refraction in two groups of patients implanted with standard biconvex IOLs (black symbols) and an anterior meniscus IOL (blue symbols). From reference [9].

**Concluding Remarks**

Peripheral optics and vision in the pseudophakic eye have been largely ignored. However, the potential impact in the quality of vision is quite significant. New designs of IOLs optimizing peripheral optics are already underway. Alternative design options and new testing strategies should be developed in the future to incorporate these ideas into the mainstream cataract practice.

## 18. Advances in contact lens optics

Pete Kollbaum, OD, PhD
Indiana University, 800 East Atwater Avenue, Bloomington, IN 47405, USA

[ kollbaum@iu.edu ]

**Status**

Although the exact time is not precisely known, the earliest form of contact lenses were described by the early 1500's. These (dangerous) glass lenses remained largely unchanged until the mid 1900's when polymethylmethacrylate (PMMA) plastic lenses were first introduced. Like the earlier lenses, however, these lenses remained comfortable and safely wearable for only short periods of time as they prevented the necessary oxygen from reaching the cornea. Fortunately, only a few years later "soft" hydrophilic (polymacon) lenses were developed. These original lenses aimed to correct the refractive error of the eye. They did this in the simplest form by changing the optical path length of the eye plus lens. Historically, these lenses corrected spherical and regular/irregular (e.g., astigmatic) refractive errors, but as society has progressed there has been significant interest in these lenses accomplishing other and more sophisticated goals. Specifically, the goals of contact lenses have expanded optically so the lenses aid eyes and correct vision better and in different ways, but the goals of lenses have also grown beyond merely those related to correcting vision. This paper briefly discusses a few optical or optically related advancements of contact lens corrections, as well as associated technological advancements necessary to achieve these advancements. Discussion of non-optical advancements, such as those related to lens materials (e.g. drug eluting) or non-optical features can be found elsewhere.

**Current and Future Challenges**

On average, human eyes contain approximately +0.15 microns (6 mm pupil) of primary spherical aberration [1]. All spherical lenses also contain inherent levels of spherical aberration which vary as a function of lens power. To counteract these two issues, contact lenses with aspheric surfaces are commonly employed [2]. Although some lenses have been manufactured aiming to introduce no spherical aberration regardless of lens power, none are currently commercially available. Many lenses are made with spherical aberration control to introduce an amount of spherical aberration approximately equal in magnitude but opposite in sign to that of the average eye (during relaxed accommodation) to reduce the total overall aberration when on the eye. These aspheric lens surfaces now form the basis for many types of lenses beyond those which correct myopia and hyperopia.
To correct non-rotationally symmetric optical errors, such as astigmatism, lenses must be stabilized. Stabilization methods include prism ballast, peri-ballast or modified prism ballast, and dual thin zone. These mechanisms may also induce asymmetric aberrations, however. Non-rotationally symmetric, stabilized lens designs also initially formed the basis of lenses aimed to counteract for the age-related loss of the eye's accommodative ability which happens during presbyopia. Currently, however, rotationally symmetric designs are the most common, and often occur through the introduction of either primary/secondary spherical aberration [3] or through annular segments of different refractive powers [4]. Diffractive designs are also possible, but not currently commercially available in contact lens designs. The newest classification of multifocal lenses have become broadly termed "extended depth of focus" contact lenes. This somewhat ambiguous term could theoretically apply to many lens designs, but generally contains lenses which have non-monotonic, non-aspheric, and/or aperiodic profiles. These lenses often contain multiple higher order spherical aberrations, conical surfaces (e.g., axicons) [5], or unique asymmetric angular elements [6,7].

Multifocal contact lens designs have also been recently used to slow myopia progression in children, based on the theory that myopia progression is due to peripheral hyperopic focus. Specifically, contact lenses which aim to control myopia progression, therefore, aim to induce peripheral myopic defocus [8]. Most designs explored or currently utilized to control axial eye growth are somewhat similar to the rotationally symmetric contact lenses for presbyopia, such as those containing annular rings of multiple optical powers.

**Advances in Science and Technology to Meet Challenges**

All contact lenses typically decenter when on the eye relative to the pupil center [9]. Any on-eye decentration of a lens which contains aberration, such as a multifocal or myopia control design, however, will introduce other aberration in direct proportion to the amount of lens aberration and decentration [10]. To compensate for this issue, stabilized lenses with decentered optical zones (to align with the pupil center) have been explored [11].

An additional consideration of multifocal lenses, such as those used to correct presbyopia or slow eye growth is that lenses with multiple optical powers simultaneously create multiple images seen by the wearer. Although not yet incorporated into a commercialized lens design some attempt has been explored to reduce the visibility of multiple images created by the multiple refractive zones by combining specific levels of aberration within the defocus introducing zones (e.g., opposite sign spherical aberration to that of the introduced defocus, [12]. Additionally, recently, non-coaxial optical segments have been introduced in a myopia control lens design with the aim of providing myopic defocus, but defocus not focused on the optical axis, making the blur associated with this defocus less noticeable to the wearer [13,14].

An alternative way to ameliorate issues related to multiple images created by multiple lens zones or powers would be through lenses which proactively change their optical power, such as through electroactive lenses. Electroactive spectacles designed with a liquid crystal between the front and back lens surfaces were introduced in the early 2000's, but contact lens technology is now being explored [15]. Lens design, electrode material choices, communication, and power sources are key hurdles to address [16]. Several mechanisms to drive power change have been explored, including such novel ways as using electrooculographic signals from blinking to switch the display from a distance to near vision correcting optical power [15].

Electroactive lenses could be considered a subset of 'smart' contact lenses, which integrate active processes. Other smart lenses have been developed for diagnosis or treatment of disease, such as to monitor eye pressure [17], blood glucose [18], cancer [19], etc. They have also been developed to incorporate eye tracking or wireless eye-machine interaction [20]. Related, attempts with some reported early successes have been made to incorporate augmented reality into a soft contact lens (e.g., Mojo vision). However, no products have been fully realized or are yet commercially available.

**Concluding Remarks**

Advances in manufacturing technology aim to keep pace with the increased demand for optical optimizations to enhance vision. These necessary technological advances also allow current biomedical technologies to accomplish more than their originally intended or singular functions and better improve condition or disease diagnosis, management, and treatment.

**Acknowledgements**

Kim Jedlicka, OD

## 19. Neurosciences applied to vision health

Stéphanie C. Thébault

Laboratorio de Investigación Traslacional en Salud Visual (D-13), Instituto de Neurobiología, Universidad Nacional Autónoma de México (UNAM), Querétaro, Mexico

[ sthebault@comunidad.unam.mx ]

**Status**

Visual science is the model system of neuroscience, not only because visual paradigms are widely used to understand behavior and cognitive functions, but also because its results are relevant to all other fields, including molecular and cellular biology, cognitive science, ophthalmology, psychology, computer science, optometry, and education. The reader can consult the more than 100 book chapters that cover the historical foundations to the latest findings in visual neuroscience [1], a recent revision article in which among the nearly 40,000 publications in this field (almost 15,000 in the last 5 years), the 25 articles that have contributed most to the advancement of visual neurosciences have been reviewed [2], as well as the original works for more details.

A deeper understanding of the early wiring of the visual system, how the brain generates vision, and the effect of visual deprivation and plasticity will help translate basic research into effective approaches and treatments to prevent and repair visual impairment and dysfunction. Visual deficiencies are characterized by a decrease in visual acuity and/or a restriction of the visual field. They have several stages, the most serious being blindness, and various causes. Glaucoma, macular degeneration, amblyopia ("lazy eye"), myopia, are among the most notable types, while retinitis pigmentosa (RP) is the most common type of inherited retinal dystrophy.

Discoveries in visual neuroscience largely follow advances in microscopy, genetics, physiology, and data processing. For example, human genetic studies, the biochemical characterization of rhodopsin, the first G-protein-coupled receptor identified and the molecular substrate of phototransduction, the deciphering of retinal circuits combined with optogenetic therapy that transforms surviving cells into artificial photoreceptors, have recently enabled partial recovery of vision in a patient with RP [3].

Artificial intelligence-based models have become the tools of choice to predict the neural responses of any brain area, including the visual cortex [4], to any possible input stimulus, and to predict and detect neuro(retino)pathies based on retina and optic nerve head imaging data. The nexus between retinal function and the brain has been very close lately because psychiatric disorders resulting from an imbalance in neurotransmitters, might be more stabilized if the optic nerves are sending balanced signals into the brain [5] and might be diagnosed thanks to visual system information [6]. Similarly, vision therapy for concussion-related vision disorders, strokes of the optic nerve or brain, and others benefits from incorporating biofeedback of the eye-brain connection [7].

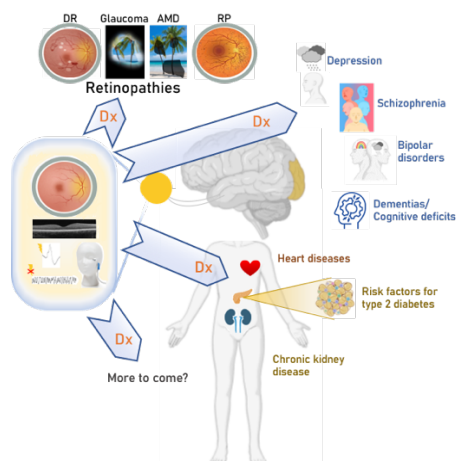

**Figure 1.** Imaging of the retinal head and optic nerve and electrophysiological data are the basis for the diagnosis (Dx) of the main retinopathies, including diabetic retinopathy (DR), age-related macular degeneration (AMD) and retinitis pigmentosa (RP), but are also currently being studied extensively in the context of the differential diagnosis of psychiatric pathologies. These data are also used to predict the risk of cardiovascular, metabolic and kidney diseases.

**Current and Future Challenges**

The main lines of research of neurosciences applied to visual health are visual prostheses and the adapted use of natural strategies for the (re)generation of nervous tissue involved in vision and its function to recover vision loss associated with RP, macular degeneration, glaucoma, amblyopia [8] or other traumatic injuries. In addition to technological advances, these approaches require knowledge of the neural visual code and the early wiring mechanisms through which visual system is refined.

Detailed characterization of the light responses of each retinal cell type to diverse visual stimuli has begun, but their diversity is staggering. The situation is somewhat similar within the thalamus and cortex. The processing of low-level features like shape, curvature, contrast, and motion at the retinal level and its use by the brain remains to be understood to uncover coding principles of visual perception. Particularly, much remains to be clarified regarding how the perception (ventral) and action (dorsal) pathways interact, how feedback circuits modify these feedforward pathways, and to what extent such computations are distributed across visual hierarchy levels. Furthermore, artificial vision requires knowing how this code is altered in blind individuals and the effect of visual deprivation and plasticity [9], as well as adapting technologies that enable to read–write neuronal activity in vivo, at single-cell resolution [10,11,12], to manipulate neuronal ensemble activity. Solutions for enlarging the visual field, increasing the visual field size, increasing spatial and temporal resolution of visual prostheses, and improving electrode–tissue interface are also needed [9].

An additional contribution of bioelectronic medicine is to complement neural visual code models with genetic, physiological, clinical, and environmental data specific to each individual to develop tailor-made therapeutic interventions [13]. In addition to data privacy and security measures, a medium- to short-term challenge is to mitigate biases due to the sampled population and the technologies used for data collection [13].

In parallel, correcting mutations in the phototransduction machinery with gene therapy and genome-editing tools is a current challenge to interrupt or reverse many blinding disorders [14]. Stem cell therapy treatment strategy also shows promise in effectively improving vision loss related to macular degeneration [15].

Models of visual system function will also need to integrate other cell types than neurons, such as endothelial cells and neuroglia. They both play a leading role in neuromodulation, which encompasses the homeostatic support and immune defense of nervous tissue that have a direct impact on the pathogenesis of ocular neurodegenerative diseases, such as glaucoma, macular degeneration, and diabetic retinopathy [13,16].

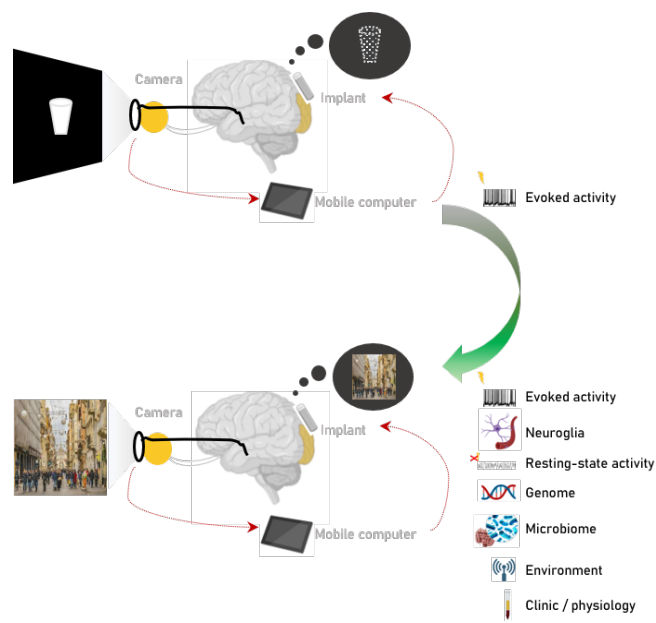

**Figure 2.** The degree of vision recovery in cases of inherited retinal dystrophies made possible by current visual prostheses (top scenario) should be significantly improved by complementing stimuli-evoked patterns of neuronal activity with data on neuroglial function, basal neuronal activity, genetic, clinical, physiological and environmental data, including the microbiome (bottom scenario).

**Advances in Science and Technology to Meet Challenges**

Several research and technological opportunities can be seen in the field of neuroscience applied to visual health. The main ones are common to any neuroscience.

With the current availability of recordings of more than 1,000 neurons at a time, the difficulty of analyzing these population data arises, the current analysis pipelines lead only to activity correlation. In addition to an easy access to a huge capacity for storing and processing information, there is therefore a growing need to develop programs and markers that evidence causal links, so that we might understand how population codes relate to distinct cell types and cell type-specific circuits and, subsequently or in parallel, to decipher how all this neural code is decoded and converted into visual perception [17]. In addition, current techniques do not yet allow for unit recordings during visual perception tasks. Some responses will have to wait for the development of time-specific retinal neuronal activation or inhibition by sonogenetics [18] for example, because in the presence of functional photoreceptors, there is no way to apply optogenetics. The picture will also be complete when field and unit recordings and analysis will take into account intrinsic neuronal activity, also known as noise.

Once this decoding under physiological conditions will be complete, merging this knowledge with predictive models that closely monitor the evolution of visual dysfunction and visual impairment will likely offer invaluable insights in identifying key points of intervention and tailoring therapeutic approaches to counteract or slow the progression of these conditions. Real-time monitoring using wearable sensors and implantable devices is one of the advancements needed to provide continuous health data. Furthermore, computational models of human vision will be of huge help to the resolution of these goals, but we cannot forget that these are powered with organic data coming from humans or clinically relevant models.

**Concluding Remarks**

One of the most striking facts about neurosciences applied to visual health is that the eye is at the forefront of developing therapies for genetic diseases [3].
Visual neurosciences contribute to visual and other organ (brain, heart, kidney) health. The eye-brain connection [7] will likely play a more prominent role in brain rehabilitation in the near future thanks to the growing emergence of neuro-optometry [19].
It is generally said that the collaborative work of many disciplines is the way forward to deepen our understanding of the visual system function and for the use of the knowledge gained to understand the mechanisms of the disease and develop treatments to pave the way for the rapid diagnosis and repair of visual impairment, visual dysfunction, and blindness. Nevertheless, faced with the ocean of data and the neglect of certain lines of research [2], I come to wonder if the achievement of these noble goals will not also depend on not losing course and devoting time to the coordinated collaborative work of the people who make the disciplines (ophthalmologists vs. ophthalmology, neuroscientists vs. neuroscience, biomedical engineers vs. engineering, etc.).
Anyway, visual neuroscience has a long way to go and a bright future.

**Acknowledgements**

This research was supported by the UNAM-DGAPA PAPIIT grants IN212823 and the National Council of Humanities, Science and Technology of Mexico (CONAHCYT) CF-2019-1759.